\documentclass[twocolumn]{aastex631}

\usepackage{color}
\usepackage{CJK}
\usepackage{graphicx}
\usepackage{amsmath}
\usepackage{amssymb}
\usepackage{appendix}

\usepackage[mathlines]{lineno}

\hypersetup{pdfauthor={Name}}

\graphicspath{{./}{figures/}}

\shorttitle{FrankenBlast Supernova Hosts}
\shortauthors{Nugent et al.}

\begin{document}

\title{Characterizing Supernova Host Galaxies with \texttt{FrankenBlast}: A Scalable Tool for Transient Host Galaxy Association, Photometry, and Stellar Population Modeling} 

\correspondingauthor{A. E. Nugent}
\email{anya.nugent@cfa.harvard.edu}

\newcommand{\NU}{\affiliation{Center for Interdisciplinary Exploration and Research in Astrophysics (CIERA) and Department of Physics and Astronomy, Northwestern University, Evanston, IL 60208, USA}}

\newcommand{\Purdue}{\affiliation{Purdue University, 
Department of Physics and Astronomy, 525 Northwestern Avenue, West Lafayette, IN 47907, USA}}

\newcommand{\CfA}{\affiliation{Center for Astrophysics\:$|$\:Harvard \& Smithsonian, 60 Garden St. Cambridge, MA 02138, USA}}

\newcommand{\UCSC}{\affiliation{Department of Astronomy and Astrophysics, University of California, Santa Cruz, CA 95064, USA}}

\newcommand{\IS}{\affiliation{Centre for Astrophysics and Cosmology, Science Institute, University of Iceland, Dunhagi 5, 107 Reykjav\'ik, Iceland}}

\newcommand{\DAWN}{\affiliation{Cosmic Dawn Center (DAWN), Niels Bohr Institute, University of Copenhagen, Jagtvej 128, 2100 Copenhagen \O, Denmark}}

\newcommand{\PUCV}{\affiliation{Instituto de F\'isica, Pontificia Universidad Cat\'olica de Valpara\'iso, Casilla 4059, Valpara\'iso, Chile}}

\newcommand{\IPMU}{\affiliation{Kavli Institute for the Physics and Mathematics of the Universe (Kavli IPMU), 5-1-5 Kashiwanoha, Kashiwa, 277-8583, Japan}}

\newcommand{\PSU}{\affiliation{Department of Astronomy \& Astrophysics, The Pennsylvania State University, University Park, PA 16802, USA}}

\newcommand{\ICDS}{\affiliation{Institute for Computational \& Data Sciences, The Pennsylvania State University, University Park, PA, USA}}

\newcommand{\IGC}{\affiliation{Institute for Gravitation and the Cosmos, The Pennsylvania State University, University Park, PA 16802, USA}}

\newcommand{\Swin}{\affiliation{ Centre for Astrophysics and Supercomputing, Swinburne University of Technology, Hawthorn, VIC, 3122, Australia}}

\newcommand{\Curtin}{\affiliation{ International Centre for Radio Astronomy Research, Curtin University, Bentley, WA 6102, Australia}}

\newcommand{\MQ}{\affiliation{Department of Physics \& Astronomy, Macquarie University, NSW 2109, Australia}}

\newcommand{\MQAAAstro}{\affiliation{Macquarie University Research Centre for Astronomy, Astrophysics \& Astrophotonics, Sydney, NSW 2109, Australia}}

\newcommand{\CSIRO}{\affiliation{CSIRO, Space and Astronomy, PO Box 76, Epping NSW 1710 Australia}}

\newcommand{\KICP}{\affiliation{Kavli Institute for Cosmological Physics, The University of Chicago, 5640 South Ellis Avenue, Chicago, IL 60637, USA}}

\newcommand{\UChicago}{\affiliation{Department of Astronomy \& Astrophysics, University of Chicago, 5640 S Ellis Avenue, Chicago, IL 60637, USA}}

\newcommand{\UA}{\affiliation{University of Arizona, Steward Observatory, 933~N.~Cherry~Ave., Tucson, AZ 85721, USA}}

\newcommand{\EFI}{\affiliation{Enrico Fermi Institute, The University of Chicago, 933 East 56th Street, Chicago, IL 60637, USA}}

\newcommand{\mpia}{\affiliation{Max-Planck-Institut f\"{u}r Astronomie (MPIA), K\"{o}nigstuhl 17, 69117 Heidelberg, Germany}}

\newcommand{\GWU}{\affiliation{Department of Physics, The George Washington University, Washington, DC 20052, USA}}

\newcommand{\UCB}{\affiliation{Department of Astronomy, University of California, Berkeley, CA 94720-3411, USA}}

\newcommand{\RU}{\affiliation{Department of Astrophysics/IMAPP, Radboud University, PO Box 9010,
6500 GL, The Netherlands}}

\newcommand{\LJMU}{\affiliation{Astrophysics Research Institute, Liverpool John Moores University, IC2, Liverpool Science Park, 146 Brownlow Hill, Liverpool L3 5RF, UK}}

\newcommand{\LU}{\affiliation{School of Physics and Astronomy, University of Leicester, University Road, Leicester. LE1 7RH, UK}}

\newcommand{\Adler}{\affiliation{The Adler Planetarium, 1300 South DuSable Lake Shore Drive, Chicago, IL 60605, USA}}

\newcommand{\ANU}{\affiliation{Research School of Astronomy and Astrophysics, Australian National University, Canberra, ACT 2611, Australia}}

\newcommand{\Car}{\affiliation{Cardiff Hub for Astrophysics Research and Technology, School of Physics \& Astronomy, Cardiff University, Queen's Buildings, Cardiff CF24 3AA, UK}}

\newcommand{\IAIFI}{\affiliation{The NSF AI Institute for Artificial Intelligence and Fundamental Interactions}}

\newcommand{\MIT}{\affiliation{Department of Physics and Kavli Institute for Astrophysics and Space Research, Massachusetts Institute of Technology, 77 Massachusetts Avenue, Cambridge, MA 02139, USA}}

\newcommand{\Hawaii}{\affiliation{Institute for Astronomy, University of Hawai‘i, 640 N. A‘ohoku Pl., Hilo, HI 96720, USA}}

\newcommand{\Weizmann}{\affiliation{Department of Particle Physics and Astrophysics, Weizmann Institute of Science, 234 Herzl St, 7610001 Rehovot, Israel}}

\newcommand{\Minnesota}{\affiliation{School of Physics and Astronomy, University of Minnesota, Minneapolis, MN 55455, USA}}
\author[0000-0002-2028-9329]{Anya E. Nugent}
\CfA

\author[0000-0002-5814-4061]{V.~Ashley Villar}
\CfA
\IAIFI

\author[0000-0003-4906-8447]{Alex Gagliano}
\IAIFI
\CfA
\MIT

\author[0000-0002-6230-0151]{David O. Jones}
\Hawaii

\author[0009-0000-6748-4319]{Asaf Horowicz}
\Weizmann

\author[0000-0002-9886-2834]{Kaylee de~Soto}
\CfA

\author[0000-0001-9269-5046]{Bingjie Wang}
\thanks{NHFP Hubble Fellow}
\affiliation{Department of Astrophysical Sciences, Princeton University, Princeton, NJ 08544, USA}

\author[0000-0001-8405-2649]{Ben Margalit}
\Minnesota

\begin{abstract}
We present \texttt{FrankenBlast}, a customized and improved version of the \texttt{Blast} web application. \texttt{FrankenBlast} associates transients to their host galaxies, performs host photometry, and runs a innovative SED fitting code to constrain host stellar population properties--all within minutes per object. We test \texttt{FrankenBlast} on 14,432 supernovae (SNe), $\approx$half of which are spectroscopically-classified, and are able to constrain host properties for 9262 events. When contrasting the host stellar masses ($M_*$), specific star formation rates (sSFR), and host dust extinction ($A_V$) between spectroscopically and photometrically-classified SNe~Ia, Ib/c, II, and IIn, we determine that deviations in these distributions are primarily due to misclassified events contaminating the photometrically-classified sample. We further show that the higher redshifts of the photometrically-classified sample also force their $M_*$ and sSFR distributions to deviate from those of the spectroscopically-classified sample, as these properties are redshift-dependent. We compare host properties between spectroscopically-classified SN populations and determine if they primarily trace $M_*$ or SFR. We find that all SN populations seem to both depend on $M_*$ and SFR, with SNe II and IIn somewhat more SFR-dependent than SNe Ia and Ib/c, and SNe Ia more $M_*$-dependent than all other classes. We find the difference in the SNe Ib/c and II hosts the most intriguing and speculate that SNe Ib/c must be more dependent on higher $M_*$ and more evolved environments for the right conditions for progenitor formation. All data products and \texttt{FrankenBlast} are publicly available, along with a developing \texttt{FrankenBlast} version intended for Rubin Observatory science products.
\end{abstract}

\keywords{supernovae, galaxies, surveys, stellar populations}

\section{Introduction}
\label{sec:intro}
The host galaxies of astrophysical transients give crucial insight on their progenitors and formation pathways. For example, the presence of Type Ia supernovae (SNe; \citealt{mannucci2005, mm2012, maoz2014, wang2013, Pan2014, chen2021, wiseman2021}) and short $\gamma$-ray bursts (GRBs; \citealt{leibler2010, fong2013, fong2022, nugent2022, jeong2024}) in quiescent galaxies and passive regions within their hosts first illustrated that their progenitors have long delay times and must arise from compact object origins. Indeed, SNe Ia are thought to originate from the thermonuclear detonation of a white dwarf (WD) that approaches the Chandrasekhar mass limit (see \citealt{maoz2014} for a review), whereas short GRBs ($\gamma$-ray duration $\lesssim 2$~sec) have been connected to the population of gravitationally-wave detected neutron star mergers \citep{aaloc+17, gvb+17, sfk+2017}. 

The intrinsic connection of Type Ib/c SNe, Type II SNe, Type IIn SNe, and long GRBs to actively star-forming hosts \citep{prieto2008, anderson2009, arcavi2010, svensson2010, li2011, anderson2012, sanders2012, Perley2013, taddia2015, vergani2015, anderson2016, graur2017, schulze2021,  taggart2021, ransome2022, qz2024} is owed to their young, massive star progenitors that undergo core-collapse \citep{smartt2015}. Their SN spectroscopic signatures \citep{filipenko1997} and subtle differences in their host populations highlight that they arise from distinct end stages of massive star evolution. In particular, the lack of hydrogen lines in SN Ib/c spectra suggest that their progenitors were ``stripped" of their outer hydrogen envelope, while SN II and IIn spectra contain hydrogen lines, implying their massive star progenitors have still retained this outer layer. SNe IIn have narrow spectral features \citep{schlegel1990, ransome2021}, rather than the broad-lines observed in SNe II(P/L) spectra, which are likely caused by strong circumstellar medium (CSM) interactions. The hosts of SNe Ib/c appear to be more massive and metal-rich than SNe II and IIn hosts \citep{prieto2008,arcavi2010,graur2017, schulze2021, qz2024}, insinuating that stripped-envelope SNe may form in distinct environmental conditions.  

Another kind of SN, hydrogen-poor super-luminous SNe (SLSNe-I), are almost exclusively observed in star-forming, low-metallicity dwarf galaxies, suggesting they too derive from massive star explosions \citep{chen2013, lunnan2014, lunnan2015, angus2016, perley2016,schulze2018,schulze2021}. Unlike other core-collapse SNe (CCSNe), SLSNe-I require a central engine to power their high peak luminosities and timescales, and, thus, a young, rapidly rotating magnetar model has been proposed for their origins \citep{nicholl2017,blanchard2020,hsu2021}. Moreover, they occur in regions within their hosts with less fractional flux than other CCSNe, hinting that their progenitors follow a distinct massive star evolutionary path \citep{hsu2024}. Finally, the small population of luminous fast blue optical transients (LFBOTs) seemingly prefer dwarf hosts with a range of global and local star-formation activity, making it difficult to explain their origins with either a massive star or compact object progenitor \citep{cmt+2020, hpk+2020, perley2021, sms23,chrimes2024}. Building up larger populations of LFBOTs and their hosts may eventually reveal which progenitor model is preferred.

Understanding transient host populations and distinguishing the host environments between transient populations will be imperative for characterizing the enormous transient populations (and their progenitors) discovered by upcoming missions such as the Vera C. Rubin Observatory's Legacy Survey of Space and Time (Rubin; \citealt{LSST, lsst2019}) and the \textit{Nancy Grace Roman Space Telescope} (\textit{Roman}; \citealt{spergel2015}). The majority of events observed through both missions will not be spectroscopically classified given current and future resources, necessitating that we instead rely on properties independent of their spectroscopic features to classify them. A plethora of transient classifiers have already been built to classify transients solely on their lightcurves \citep{muthukrishna2019,villar2019, Hosseinzadeh2020, moller2020, villar2020,  boone2021, sanchez2021, qu2022, deSoto2024, btsbot2024}, with several studies finding that photometric classification (particularly at early times) improves when including transient host galaxy information \citep{foley2013, jones2017, gomez2020,gomez2023,gomez2023b,gagliano2023, kisley2023,sheng2024,villar2024impact,boesky2025}. Importantly, this highlights that transient populations can be studied in the absence of the specific properties of transients themselves via their hosts. However, this does have the potential to bias transient samples; thus, careful attention is required when employing host galaxies for classification.

Despite the utility of transient host galaxies in helping to characterize them, analyzing large galaxy populations has generally been quite challenging. Constraining galaxy stellar population properties quickly and in a computationally inexpensive way, which is often desired for large galaxy population studies, has often been done through matching observed spectral energy distributions (SEDs) to templates of elliptical and spiral galaxies or employing simple stellar population (SSP) models. Unfortunately, both of these methods are prone to severely misrepresenting the true properties of observed galaxies \citep{Conroy2013}. Traditional Bayesian-inference based SED fitting techniques, which can handle more complex, composite stellar population models that obtain more realistic constraints on properties than the aforementioned methods, can be quite computationally expensive, with some codes taking $\gtrsim1$~day (on one CPU) to fit a single galaxy \citep{BEAGLE,BAGPIPES,jlc+2021}. Thus, using these methods in real-time with Rubin and \textit{Roman} detected transients would require an astronomical amount of computational resources. 

Recent developments in SED fitting have alleviated some of these issues. Taking advantage of machine learning mechanisms, \citet{wlv+2023} developed a galaxy SED fitting technique that is 500 times faster than the Bayesian-inference approach, while still maintaining their more complex stellar population models and realistic posteriors on various stellar population properties. This technique has already been incorporated into the \texttt{Blast} web application\footnote{https://blast.scimma.org/} \citep{Blast}, a live application for determining host galaxies and rapidly deriving stellar population properties for recently reported events on the Transient Name Server (TNS). {\tt Blast} has been used for host galaxy analyses in a number of recent publications, including: \citet{AbreuPaniagua2025,Gagliano2025,Hayes2025,Hoogendam2025a,Hoogendam2025b,Lebaron2025,Pessi2025}.  Capitalizing on tools such as \texttt{Blast}---those that focus on computational speed to enable large, uniform studies---will be vital for any host or transient-related science performed during the Rubin and \textit{Roman} eras.

Here, we develop a customized version of \texttt{Blast}, which we call \texttt{FrankenBlast}\footnote{https://github.com/anugent96/frankenblast-host} (Zenodo: \dataset[doi:10.5281/zenodo.17555070]{https://doi.org/10.5281/zenodo.17555070}), that associates transients to their host galaxies, performs aperture photometry on the hosts using archival data from an array of public surveys, and determines host galaxy stellar population properties \textit{all within minutes per object}. While \texttt{Blast} is a web-application that publicizes all host associations and host stellar population properties using a set pipeline, \texttt{FrankenBlast} is a package that users can download locally and manipulate for their own scientific needs. We furthermore envision expanding \texttt{FrankenBlast} to include pipelines specific to upcoming surveys, such as Rubin (already implemented for commissioning data) and \textit{Roman}. In addition, \texttt{FrankenBlast} notably updates and improves the host association method originally employed in \texttt{Blast}, as well as the SED fitting tool, which now includes a photometric redshift estimator. Future versions of \texttt{Blast} will include these updates. In this study, we also use \texttt{FrankenBlast} to determine the hosts of 6676 spectroscopically-classified and 7756 photometrically-classified SNe and study their host stellar population properties. 

In Section \ref{sec:sample}, we present our transient sample. We present our host association method, determine hosts for our transient sample, and compare our host associations to those previously determined for a subset of our sample in Section \ref{sec:host}. We outline our methods for performing host photometry in Section \ref{sec:phot}. In Section \ref{sec:sp_model}, we discuss our stellar population modeling technique, the implementation of this technique on our transient host sample, and a comparison of this technique with another SED fitting code. We compare the host properties of our spectroscopically and photometrically-classified samples in Section \ref{sec:specvphot}. In Section \ref{sec:transientcompare}, we compare the host properties between different SN classes and compare our results to previous literature. We interpret these results in the context of SN progenitors and discuss our work's applicability to Rubin-era science in Section \ref{sec:disc}. Finally, we list our main conclusions in Section \ref{sec:conc}. All data products from this study are available on Zenodo: \dataset[doi:10.5281/zenodo.16953206]{https://doi.org/10.5281/zenodo.16953206}.

Unless otherwise stated, all observations are reported in the AB magnitude system and have been corrected for Galactic extinction in the direction of the object of interest using the \citet{sf11} extinction maps.  We employ a standard WMAP9 cosmology of $H_{0}$ = 69.6~km~s$^{-1}$~Mpc$^{-1}$, $\Omega_\textrm{m}$ = 0.286, $\Omega_\textrm{vac}$ = 0.714 \citep{Hinshaw2013, blw+14}. 

\section{Host Galaxy Sample}
\label{sec:sample}

\begin{figure*}
\centering
\includegraphics[width=1.0\textwidth]{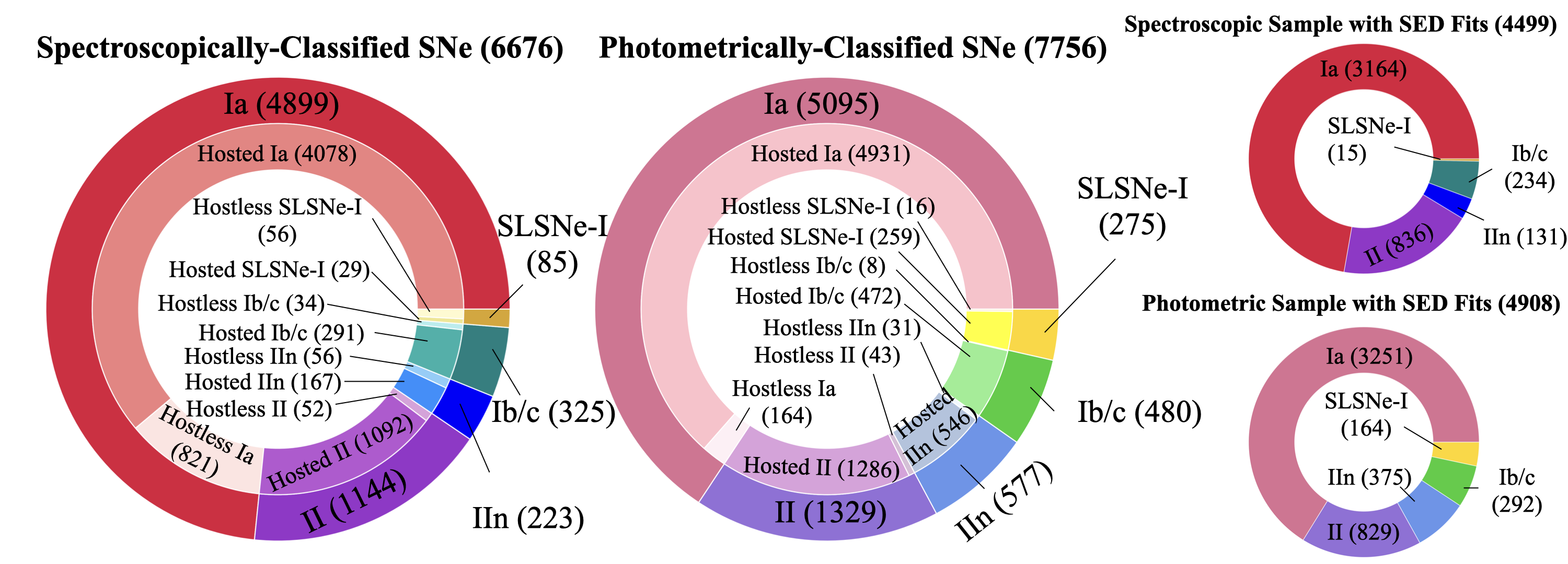}
\vspace{-0.1in}
\caption{The breakdown of our spectroscopically and photometrically classified SN samples. \textit{Left:} The number of SNe Ia, SNe II, SNe IIn, SNe Ib/c, and SLSNe-I in our spectroscopically-classified sample (outer circle), and the number of events, separated by class, with and without \texttt{Pröst} host associations (inner circle). We claim that the events without host associations are ``hostless". \textit{Middle:} The same, but for the photometrically-classified SN sample. \textit{Right:} The number of spectroscopically (top) versus photometrically (bottom) classified events with host SED fits through \texttt{SBI++}.}
\label{fig:sample}
\end{figure*}

We perform our host galaxy analysis on the transient sample analyzed in \citet{deSoto2024} and the Young Supernova Experiment Data Release 1 (YSE DR 1) SN sample \citep{jones2021, ams+2023}. \citet{deSoto2024} query light curves of transients detected between 2018-2023 on TNS. All transients in their sample were observed with the Zwicky Transient Facility (ZTF; \citealt{ZTF}), a wide-field transient survey located at the Palomar Observatory. A large portion of their spectroscopic sample was also observed as part of the ZTF Bright Transient Survey {BTS; \citealt{Fremling2020}), which aims to spectroscopically classify all ZTF transients with peak magnitude $<19$~mag. We supplement this sample with ZTF transients detected until January 2025, using the same sample selection methods outlined in \citet{deSoto2024}. The spectroscopically-classified SN sample (excluding non-SN transients like cataclysmic variables, active galactic nuclei, tidal disruption events, luminous blue variables, luminous red novae, etc., and hydrogen-rich SLSNe) contains 6199 events out to $z\approx0.6$, with median redshift $z=0.058$. We find that 73\% of the spectroscopically classified sample are SNe Ia, 17\% are SNe II, 3\% are SNe IIn, 4.6\% are SNe Ib/c, and 1.3\% are SLSNe-I. 

The photometric sample comprises an additional 6283 SN-like transients. Photometric classifications for this sample are achieved through the SN classification pipeline, \texttt{Superphot+} \citep{deSoto2024}, which achieves state-of-the-art performance on ZTF light curves. \texttt{Superphot+} fits a parametric model to the $g$- and $r$-band ZTF lightcurves of the sample to extract features from the full SN sample. The spectroscopic subsample is then used to train a Gradient Boosting Machine (GBM) classifier, which is applied to the remaining (photometric) set. \texttt{Superphot+} classifies transients as SNe Ia, SNe II, SNe IIn, SNe Ib/c, and SLSNe-I, determining classification probabilities for each of these SN classes (which all sum to one). For each transient, we select the SN class with the highest probability as its classification. For more details on \texttt{Superphot+}, we refer the reader to \citet{deSoto2024}. With these classifications, we find our updated photometric sample is 69\% SNe Ia, 16\% SNe II, 5\% SNe IIn, 7\% SNe Ib/c, and 3\% SLSNe-I.

The YSE DR1 dataset comprises 1975 transients (1962 excluding non-SN transients and SLSN-II) observed with the Panoramic Survey Telescope and Rapid Response System (Pan-STARRS; \citealt{PanSTARRS}, all with peak apparent magnitudes $\lesssim 22$~mag. Within the SN sample, 477 events are spectroscopically classified, spanning a redshift range of $0.003 < z < 0.224$, with median $z=0.065$. This sample is 66\% SNe Ia, 22\% SNe II, 3\% SNe IIn, 8\% SNe Ib/c, and $<1\%$ of SLSNe-I. The remaining 1485 SNe comprise the photometric sample. For consistency, we use \texttt{Superphot+} to classify the YSE DR1 photometric sample. All but 12 SNe in the YSE DR1 photometric sample are also detected in ZTF, thus we are able to determine photometric classifications for the majority of this dataset using the same methods as described in \citet{deSoto2024}. We find that 53\% of events are classified as SNe Ia, 20\% are SNe II, 18\% are SNe IIn, 4\% are SN Ib/c, and 5\% are SLSN-I.

We highlight the breakdown of SNe~Ia, SNe~II, and SNe~Ib/c for our complete spectroscopic and photometric samples in Figure \ref{fig:sample}. In total, we run \texttt{FrankenBlast} on 6676 spectroscopically classified SNe and 7756 photometrically classified SNe.

\section{Host Association}
\label{sec:host}

\subsection{Pröst}
Within \texttt{FrankenBlast}, we associate our SN sample to their host galaxies using the Python-based host association code \texttt{Pröst} \citep{Prost}\footnote{https://github.com/alexandergagliano/Prost}. \texttt{Pröst} allows users to condition host association on several different parameters, including the fractional offset of the transient, the absolute brightness of the galaxy, and the redshift of the system. Fractional offset is defined as the angular offset of the SN to a galaxy's catalog-reported center divided by the galaxy's directional light radius (DLR): the half-light radius of a galaxy in the direction of the SN \citep{sullivan2006,Gupta2016, GHOST}. Within \texttt{Pröst}, the user defines priors over each of these properties, which can be informed by their knowledge of the redshift distribution of the time-domain survey and the class of transient being associated. \texttt{Blast} originally used \texttt{GHOST} \citep{gne+2021} for host association, which uses a modified DLR method but does not take redshift into account or associate hosts probabilistically; thus, it is less favorable for our study\footnote{\texttt{Blast} has recently updated its host association method to \texttt{Pröst} and now is in alignment with \texttt{FrankenBlast}.}. After priors are defined, \texttt{Pröst} initiates a Monte Carlo routine to sample from the distribution of observed properties (brightness, redshift, and/or offset) for each galaxy within a given search cone centered on the transient position. \texttt{Pröst} assumes each galaxy's observed properties follow a Gaussian distribution, with the mean and the standard deviation assumed to be the catalog-level point estimate and associated error in that measurement. The posterior probability of each galaxy is then calculated as the prior probability density function (PDF) of the galaxy property multiplied by the likelihood PDF of each galaxy property. 

\texttt{Pröst}, finally, considers the possibility that the host is unobserved (e.g., that the true host is beyond the limiting magnitude of the survey in which host association is attempted, or the true host lies outside of the search radius). In this case, the code runs a Monte Carlo routine, sampling from the priors, and integrates the posterior probability of all potential unobserved host galaxies (by simulating a galaxy population beyond the bounds of the chosen search cone and beyond the magnitude limit of the chosen survey, and integrating the posterior probability that each of these unobserved galaxies is the true host). Then, the posteriors for each galaxy (all observed galaxies and un-observed galaxies) are summed and re-normalized. \texttt{Pröst} selects the scenario with the highest posterior in the largest number of Monte-Carlo draws, which approximates the modes of the true posteriors. 

We attempt host associations for our SN sample using the \textit{Galaxy List for Advanced Detector Era} (GLADE; \citealt{GLADE}), Pan-STARRS \citep{PanSTARRS}, and Dark Energy Camera Legacy Survey (DECaLS) Data Release 10 \citep{DECaLS}. Each survey contains optical photometry, a spectroscopic or photometric redshift, and shape measurements (semi-major axis, axis ratio, and position angle) for observed galaxies. We define a search radius of 300'' for all \texttt{Pröst} runs. For spectroscopically-classified SNe, we construct priors on redshift, fractional offset, and host absolute brightness. We set the redshift prior to be a uniform distribution between $0.0001 \leq z \leq 0.6$ (the maximum redshift of our spectroscopic sample). We further enforce a nominal SN redshift error of 5\% for sampling across Monte-Carlo runs. The likelihood is set by the SN redshift and is calculated as the probability of the reported value relative to a normal distribution with mean of each candidate galaxy's redshift and standard deviation of each host's redshift error. We assume fractional offset has a uniform prior between 0 and 10 (in units of the host’s DLR), with a $\Gamma$ function likelihood: $\Gamma(a)$, where $a=0.75$, which steeply decreases the likelihood of a galaxy being the true host for increasing fractional offsets of the transient. We define host galaxy brightness as an absolute magnitude ($M$) and set a wide uniform prior and likelihood between $-30<M<20$~mag. Absolute magnitudes for observed galaxies are determined using their given spectroscopic or photometric redshifts in each survey. For GLADE galaxies, we use $B$-band photometry and for Pan-STARRS and DECaLS galaxies, we use the median of $griz$. While testing \texttt{Pröst} on our photometrically-classified sample, we found our associations to be dominated any redshift prior if one was imposed. To mitigate this effect, we only use our previously defined priors and likelihoods for fractional offset and absolute magnitude (which has a wide enough range to essentially be negligible) when associating the photometric SN sample.

We run \texttt{Pröst} on our SN sample using all three surveys, and select the galaxy with the overall highest probability as the true host. Typically, the same highest probability host is found in multiple surveys, but there are a few cases where a higher probability faint host is found in DECaLS and not Pan-STARRS. We note that the probabilities determined by \texttt{Pröst} are re-normalized relative to each individual survey, and thus a comparison between probabilities in different catalogs in not necessarily trivial. However, for the purposes of this study, we find that this method works well. \texttt{Pröst} is typically able to determine host associations in $\approx$few seconds, with the latency dominated by remote catalog queries.

When applying our host association methods to our SN sample of the 14,432 events, we are able to find confident host associations for 13,111 events. We highlight the number of spectroscopically and photometrically classified transients with host associations in Figure~{\ref{fig:sample}}. We also show the number of ``hostless" events, which were not assigned a host association through \texttt{Pröst}. We note that, given the size of our sample, we do not inspect each association individually. We do, however, scrutinize the host associations of a subset of our sample, which we discuss in the following subsection. 

\subsection{Quality Checks \& Comparison to Previous Methods}
To determine how well \texttt{Pröst} associates transients to their hosts, we manually inspect the host associations of the full YSE DR1 transient sample (including non-SN transients). We note that previous host associations to this sample have been made in \citet{ams+2023}. We find confident host associations for 1897 of the 1975 (96\%) total events. For these events, it is generally obvious why \texttt{Pröst} chose a particular galaxy as the host (i.e., the transient was spatially coincident with the galaxy, the galaxy has a similar redshift estimate to that of the transient, the transient is closer to the chosen host than others in the field, or the galaxy is brighter than others nearby). We only find 11 cases where the host association is not obvious and required a manual re-association. It appears that these events occurred in galaxies with slightly more complicated structures (e.g., highly extended or bright regions of star formation near the SN location), and \texttt{Pröst} would either associate to one of these bright regions in the host or choose another galaxy due to the odd morphology. Indeed, galaxy shredding is a common issue for many surveys, especially at low redshift; thus these incorrect host associations are likely a fault of the survey rather than \texttt{Pröst}. We note that \texttt{Pröst} does include a very basic module for identifying instances of galaxy shredding. Other works, such as \cite{diteodoro2023}, have attempted to address this issue at the image level with machine-learning techniques. In the future, users should take advantage of these tools to mitigate possible effects of galaxy shredding in host associations. For the population without associations: either \texttt{Pröst} did not find a high-probability host (almost always because the true host is likely too faint), or we determined a host association was \textit{not} possible because the transient occurred in a crowded field with multiple high probability hosts or a bright star near the transient position prevented a secure host association. 

We next compare our associations to those determined in \citet{ams+2023}, which employs \texttt{GHOST} (previously used in \texttt{Blast}) and \texttt{Sherlock} \citep{Sherlock} to perform host associations for the YSE DR1 sample. Similar to \texttt{GHOST}, \texttt{Sherlock} makes associations through angular and physical separations between the transient position and Pan-STARRS galaxies. \citet{ams+2023} finds host associations for 1913 total events. For the 1861 events that both we and \citet{ams+2023} find host associations for, there are only 65 transients with different hosts. The majority of the time, we determine that this is because \texttt{Pröst} finds a fainter host near the position of the transient that is not detected in the Pan-STARRS imaging, or the transient spectroscopic redshift motivates a \texttt{Pröst} association to a more offset galaxy with a more similar photometric or spectroscopic redshift to the transient. There are a handful of cases where \citet{ams+2023} claims that the true host is closer to the position of the transient. However, we do not find any source with reasonable SNR at the same location. We further find that \texttt{Pröst} is able to make 36 host associations to events that \citet{ams+2023} claims there is no host; generally, these are just fainter galaxies near the transient position. We note that \citet{ams+2023} makes 52 host associations for events for which we do not have associations. Most of these cases fall into one of our exclusions for host association: e.g., the transient occurs in a crowded field, there is a bright star blocking the field, or the host is too faint. Finally, there are 26 events that neither we nor \citet{ams+2023} can secure a robust host association. Given that only $\approx3$\% of the host galaxies are different within the sample of 1861 events we and \citet{ams+2023} were able to make host associations for (and our obvious failure rate is $<1$\%), we find our techniques for host association are sufficient. Furthermore, \citet{Blast} finds that \texttt{GHOST} is 93\% accurate in \texttt{Blast}, thus \texttt{Pröst} is noticeably better.

\section{Photometry}
\label{sec:phot}
We perform global aperture photometry on all host galaxies using the methods outlined in \citet{Blast} for \texttt{Blast}, making slight adjustments to their approach in \texttt{FrankenBlast}. We collect images of each host field (if available) from the \textit{Galaxy Evolution Explorer} (GALEX; \citealt{Galex}), Pan-STARRS, DECaLS DR 9, the \textit{Two Micron All-Sky Survey} (2MASS; \citealt{2MASS}), and the \textit{Wide-field Infrared Survey Explorer} (WISE; \citealt{WISE}). This totals 17 possible images per object: GALEX FUV and NUV, Pan-STARRS $grizy$, DECaLS $grz$, 2MASS $JHK$, and WISE 1-4. In each image, we construct elliptical apertures using the \texttt{Astropy Photutils} package \citep{photutils}, which builds Kron apertures of $>3\sigma$ sources with at least 10 connected pixels (the requirement for a distinct source) from an image segmentation map. If the host is not detected in one of the filters, we extract an upper limit by rescaling the aperture measured in a neighboring filter by the spatial full-width half-maximum (FWHM) of the PSF of the filter with the non-detection. \texttt{Blast} uses this same method to determine all of their aperture sizes. Specifically, they \textit{only} measure an aperture in an optical band (typically, $g$-band) and re-scale the rest by the respective FWHM. While this method is not perfect for calculating the true aperture size in that image, \citet{Blast} suggest that this generally recovers the outer edges of galaxy light well. Our approach, in theory, should give less background noise, and thus higher signal-to-noise ratios (SNR) than the approach in \texttt{Blast}. 

When comparing our methods for 660 hosts in the YSE DR1 transient survey that had available photometry on \texttt{Blast},  we generally find good agreement between the photometry. For optical filters, we find a median magnitude difference between \texttt{Frankenblast} and \texttt{Blast} of -0.03 mag, for GALEX filters we find a median difference of -0.16~mag, for 2MASS filters we find a median difference of 0.02~mag, and for WISE filters we find a median difference of -0.3 mag. We find the most difference in photometric measurements between GALEX and WISE filters, where we generally obtain brighter photometric measurements than \texttt{Blast}. As noted in \citet{Blast}, the \texttt{Blast} photometric method likely does not create large enough aperture sizes in these bands to capture the full host flux, whereas our method does. Furthermore, we find that our optical photometry are very similar to the respective survey catalog photometry when comparing our photometry for 5000 hosts to PanSTARRS catalog photometry and 2000 hosts to DECaLS catalog photometry. We find a magnitude difference between \texttt{FrankenBlast} and PanSTARRS photometry $=-0.1^{+0.08}_{-0.17}$~mag (median and 68\% confidence interval) and $=-0.06^{+0.13}_{-0.07}$~mag for \texttt{FrankenBlast} and DECaLS photometry, suggesting that our photometry is consistent with these surveys and other techniques.

Finally, we do not implement methods to determine if an aperture is contaminated with another source or correct for contamination. Given the lower density of sources in the surveys used in this study, we do not believe that aperture contamination should affect the majority of our photometry. However, we do note that, in the next decade, with deep surveys like Rubin and \textit{Roman} that will observe $\approx 50$ galaxies per arcmin$^2$, deblending will become extremely relevant.

We compute uncertainties for all photometric measurements as described in \citet{Blast}. Uncertainties are assumed to be driven by Poisson-noise and are calibrated from the zeropoint acquired for each filter. WISE filters include correlated noise between pixels, and, thus, we apply a noise correction term that is dependent on the number of pixels in an aperture\footnote{We obtain noise correction terms from the \href{https://irsa.ipac.caltech.edu/data/WISE/docs/release/Prelim/expsup/sec2\_3f.html\#corrnoise}{WISE Users Guide Section II}, 3i Figures 1-4, and apply them to our uncertainty measurements using Section 3f, Equation 2.}. 

Within our sample, we are able to extract suitable photometry for host stellar population modeling for 11,153 SN hosts. The hosts for which we do not obtain photometry are generally too faint for an aperture measurement.

\begingroup
\setlength{\tabcolsep}{6pt} 
\renewcommand{\arraystretch}{1.3} 
\begin{deluxetable*}{l|ccccccc}[!t]
\tabletypesize{\footnotesize}
\tablecolumns{3}
\tablewidth{0pc}
\tablecaption{\texttt{Prospector} and \texttt{SBI++} Prior Distributions
\label{tab:prosp_priors}}
\tablehead{
\colhead {Parameter}	 &
\colhead {Definition}	 &
\colhead {Prior}	
}
\startdata
$z$ & redshift & $\mathcal{U}[0.001,1.501] $ \\
$\log(M_F/M_\odot)$ & total mass formed & $\mathcal{U}[6,12] $ \\
$\log(Z_*/Z_\odot)$ & stellar metallicity & \citet{gcb+05} relation, min=-1.98, max=0.19 \\
$\tau_{V,2}$ & optical depth of old stellar light & $\mathcal{N}[\mu=0.3, \sigma=1.0, \textrm{min}=0, \textrm{max}=4]$ \\
$\tau_{V,1}/\tau_{V,2}$ & fraction of optical depth of young to old stellar light & $\mathcal{N}[\mu=1, \sigma=0.3, \textrm{min}=0, \textrm{max}=2]$ \\
dust$\_$index & offset from \citet{calzetti2000} dust attenuation curve & $\mathcal{U}[-1.0,0.2]$ \\ 
$\tau_{\textrm{AGN}}$ & mid-IR optical depth & $\mathcal{U}[10,90]$ \\
$f_{\textrm{AGN}}$ & fraction of AGN luminosity in galaxy & $\mathcal{U}[10^{-5},2]$ \\
logsfrratios & ratio of SFR between two adjacent agebins & $\mathcal{T}[\mu=0, \sigma=0.3, \nu =2]$ \\
$q_\textrm{pah}$ & polycyclic aromatic hydrocarbon mass fraction & $\mathcal{N}[\mu=2, \sigma=2.0, \textrm{min}=0.9, \textrm{max}=1.1]$ \\ 
$\gamma_e$ & relative contribution of dust heated at a radiation field strength of  $U_\textrm{min}$ & $\mathcal{N}[\mu=-2, \sigma=1, \textrm{min}=-2.1, \textrm{max}=-1.9]$ \\
$U_\textrm{min}$ & minimum radiation field strength & $\mathcal{N}[\mu=1, \sigma=10, \textrm{min}=0.9, \textrm{max}=1.1]$
\enddata
\tablecomments{The \texttt{Prospector} sampled parameters, their definitions, and prior distributions, used to build the \texttt{SBI++} training samples (Section~\ref{sec:training}) and \texttt{Prospector} nested sampling test fits (Section~\ref{sec:sbi_tests}). $\mathcal{U}$ represents uniform distributions, $\mathcal{N}$ represents Normal distributions, and $\mathcal{T}$ represents StudentT distributions.}
\end{deluxetable*}

\section{Stellar Population Modeling}
\label{sec:sp_model}
To determine the stellar population properties of our host galaxy sample, we implement the \texttt{SBI++} spectral energy distribution (SED) fitting technique described in \citet{wlv+2023} in \texttt{FrankenBlast}, which is also incorporated in \texttt{Blast}. \texttt{SBI++} builds on typical simulated-based inference (SBI) methodologies, which uses normalizing flows, a specialized neural-network based algorithm that learns how to transform a simple probability distribution (in our case, a Gaussian) into a more complex distribution, to learn posterior density distributions directly from a simulated training dataset. The ``baseline" SBI requires that the data being modeled, in this case the galaxy photometry, matches the simulated training set exactly, i.e., misspecified models cannot be handled. Given that observed galaxies can have both missing data (from faintness or their field not being observed in a particular survey) and data much noisier than the training set data, baseline SBI is difficult to implement for our purposes. \citet{wlv+2023} addresses both of these issues by introducing techniques to handle missing and noisy data, which we detail in Section \ref{sec:run_sbi}.

\texttt{SBI++}, furthermore, has several advantages over traditional SED modeling codes, making it ideal for our study with \texttt{FrankenBlast} and future host analyses with Rubin and \textit{Roman}. As described in \citet{wlv+2023}, typical SED modeling codes, including \texttt{BAGPIPES} \citep{BAGPIPES}, \texttt{Prospector} \citep{jlc+2021}, and \texttt{BEAGLE} \citep{BEAGLE}, combine stellar population synthesis models (SPS) with Bayesian inference techniques (e.g., MCMC or nested sampling; \citealt{MCMC, skilling2004}) to build posterior distributions of the stellar population properties of interest. Model calls are inherently quite computationally expensive, thus sampling the full posterior distribution can take up to several days for a single SED fit: highly problematic for large host population studies. \texttt{SBI++} offers a solution to this computational complexity, as it replaces the full sampling routine with a neural network-based generative model. \citet{wlv+2023} found that \texttt{SBI++} fits were 500 times faster than these traditional methods ($\lesssim 15$~min with on a single CPU; we find the same result in our implementation). Moreover, \citet{wlv+2023} show that \texttt{SBI++} not only performs similarly in terms of accuracy to these these traditional SED modeling codes, but it also recovers more well-calibrated uncertainties on redshift than nested sampling techniques. For these reasons, we find that \texttt{SBI++} is preferred over these traditional stellar population modeling codes.

In the following subsections, we describe how we train the \texttt{SBI++} model, how we run \texttt{SBI++} on our host galaxy sample, and, finally, how our methods compare to traditional SED fitting codes.

\begin{figure*}
\centering
\includegraphics[width=1.0\textwidth]{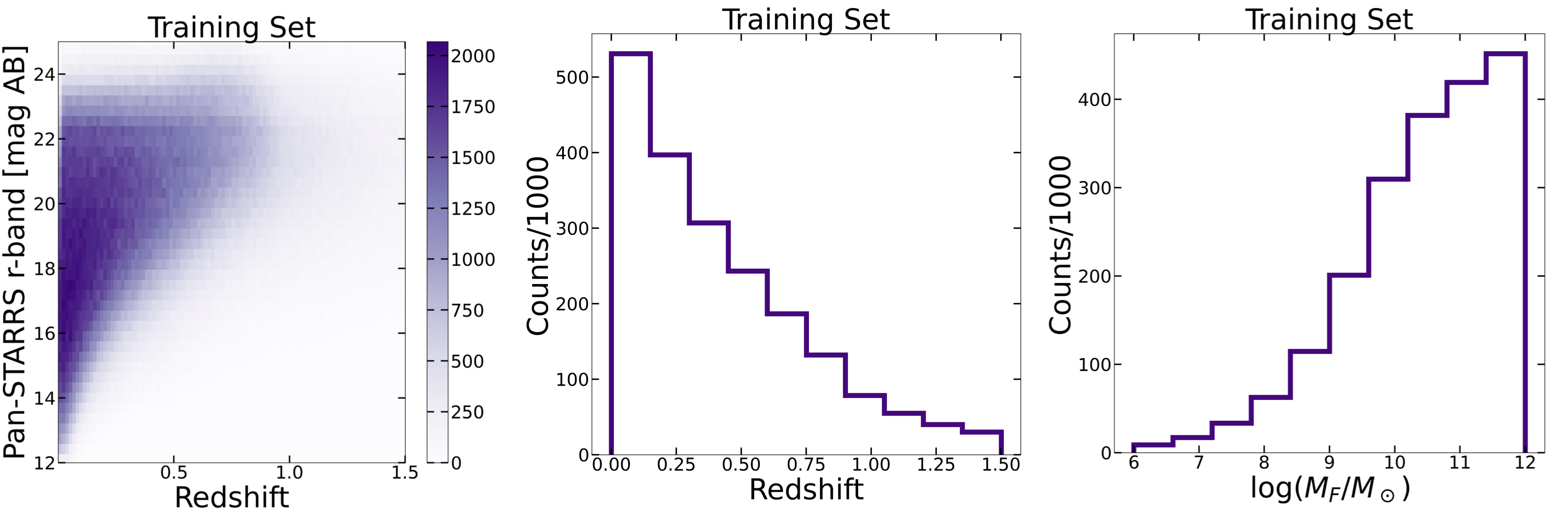}
\vspace{-0.1in}
\caption{\textit{Left:} The distribution of Pan-STARRS $r$-band magnitudes and redshifts for the mock galaxy population in our \texttt{SBI++} training set. We do not include galaxies that are much fainter or brighter than our observed host population. \textit{Middle:} The redshift distribution of the mock galaxies in the training sample. \textit{Right:} The total mass formed ($M_F$) for each galaxy in the training sample. While we use uniform priors on both redshift and $M_F$ to build the training sample, the restrictions we place on magnitude force there to be fewer high-$z$ and low-$M_F$ galaxies, which are more likely to be fainter than the detection limit of the surveys used in this study.}
\label{fig:trainingset}
\end{figure*}

\subsection{Training SBI++}
\label{sec:training}
To build a training set for our \texttt{SBI++} fitting routine, we produce mock photometry of the 17 filters discussed in Section \ref{sec:phot} from \texttt{Prospector} stellar population models, similar to the methods applied in \citet{wlv+2023} and \citet{Blast}. \texttt{Prospector} is a stellar population modeling synthesis code that employs the MIST models \citep{MIST} and MILES spectral library \citep{MILES} through \texttt{FSPS} (Flexible Stellar Population Synthesis) and \texttt{python-FSPS} \citep{FSPS_2009, FSPS_2010} to produce model photometry and spectra from a given set of stellar population properties. Our \texttt{Prospector} model includes the \citet{Chabrier2003} initial mass function (IMF), the \citet{KriekandConroy13} dust attenuation model (which applies an offset from the \citealt{calzetti2000} dust attenuation curve), and a nebular emission model \citep{bdc+2017}. We note that our nebular emission model is constrained through a gas-phase metallicity set to $Z_\odot$ as we cannot reasonably constrain it from photometry alone, given that we typically require spectral emission lines to measure it. Moreover, gas-phase metallicity is not degenerate with any other stellar population property, thus, this assumption should not affect other parameters. We further include an active galactic nuclei (AGN) torus model \citep{Leja2018}, in the event that the host has an AGN that affects their mid-IR photometry; and the \citet{DraineandLi07} IR dust emission model. We track star formation through a non-parametric \texttt{continuity} star-formation history (SFH), which assumes a constant SFR within an age bin. We include seven age bins in our SFH model: the first two range from 0 to 30 Myr and 30 to 100 Myr, and the final five are log-spaced from 100 Myr to the age of the Universe at the presumed redshift. 

We randomly sample from prior distributions of the stellar population properties of interest to generate mock photometry. We allow the following properties to be sampled: redshift, the total mass formed in the galaxy ($M_F$), stellar metallicity ($Z_*$), dust attenuated from old ($\tau_{V,2}$) stellar light and the fraction of dust attenuated from young to old stellar light ($\tau_{V,1}/\tau_{V,2}$), the offset from the \citealt{calzetti2000} dust attenuation curve, the ratio of SFR between two adjacent age bins (totaling six ratios for our seven-bin non-parametric SFH model), two AGN parameters: the fraction of possible AGN luminosity relative to the bolometric stellar luminosity ($f_\textrm{AGN}$) and a mid-IR optical depth ($\tau_\textrm{AGN}$), and three \citet{DraineandLi07} IR dust emission parameters: the polycyclic aromatic hydrocarbon mass fraction ($q_\textrm{pah}$), the contribution of dust heated ($\gamma_e$), and the minimum radiation field strength ($U_\textrm{min}$). We show all sampled parameters and their prior distributions in Table \ref{tab:prosp_priors}. We note that our model slightly differs from the one used in \texttt{Blast} (see Table 1 in \citealt{Blast}), namely in redshift where we increase the maximum from $z=0.2$ in \texttt{Blast} to $z=1.5$ in \texttt{FrankenBlast} to be better aligned with the host sample we analyze here.

Our training set comprises 2 million mock galaxies, with 2/3 of their mock photometry generated from sampling the priors in Table \ref{tab:prosp_priors} and 1/3 from uniform prior distributions on all properties except $M_F$ and $Z_*$, for which we always constrain through the \citet{gcb+05} $M_F$-$Z_*$ relation priors. We modify the \citet{gcb+05} $M_F$-$Z_*$ relation by widening the confidence intervals by a factor of two to account for any unknown potential systematics, following the guidelines in \citet{Leja2019}.  We do not include any mock galaxy in our training set with photometry well outside of the magnitude range of our observed hosts and highlight this in Figure~\ref{fig:trainingset}, although, we do note that our host population spans a wide range of magnitudes from $r=10-24$~mag. To simulate observed galaxies, we add Gaussian noise to the mock photometry using SNR determined from our host galaxy photometric measurements and photometric uncertainties from these SNRs\footnote{\texttt{FrankenBlast} uses a higher-fidelity SNR model from a much larger input data sample than \texttt{Blast}.}. To create the SNR model, we determine SNRs for all galaxies within our host population, divide the hosts into evenly-spaced magnitude bins, and calculate the mean SNR for galaxies within each bin. We impose a 1\% error floor on all observed photometry, thus this is also propagated into the SNR model. For each simulated photometric point, we interpolate over the SNR model to obtain an approximate uncertainty at the simulated magnitude. We then randomly generate a noised-up photometric point from a Gaussian distribution, using the simulated photometric point as the mean and the uncertainty as the standard deviation. We emphasize to the reader that this is a crucial step: the training set must be generated with a \textit{noise} model similar to that of the observed galaxy population in order to correctly model uncertainty on the stellar population property posterior distributions. For this reason, we do not include galaxies that have photometry determined through other methods (i.e., a larger field galaxy population from the surveys used in this study) within our noise model.

Finally, we train the model using a sequential neural posterior density estimator from the \texttt{sbi} Python package \citep{gnm+2019,tbd+2020}, following the instructions in \citet{wlv+2023}. We train two separate models, using the same training set: one with redshift as an observed property of the galaxy (to be used when there are spectroscopic redshifts or well-constrained photometric redshifts), similar to \citet{hm2022}, and one without redshift as an observed property (to be used for a photometric redshift estimate). Hereafter, we refer to these \texttt{SBI++} models as the spec-$z$ model and the photo-$z$ model. We note that the original version of \texttt{Blast} did not contain a photo-$z$ model. Here, however, we find it advantageous to fit for redshift in the model as: (i) redshift can have a large effect on the model SEDs and other stellar population property estimates, and (ii) when redshifts are unknown, having a framework where redshift is inferred simultaneously with the rest of the stellar population properties means that the uncertainties in redshift can be properly propagated into estimates of the other properties (discussed further in \citealt{wang2023}). Training the spec-$z$ model takes $\approx 2$ days and training the photo-$z$ model takes $\approx 5$ days on the Harvard Faculty of Arts and Sciences Research Computing (FASRC) Cannon cluster using one CPU.

\begin{figure*}[t]
\centering
\includegraphics[width=0.47\textwidth]{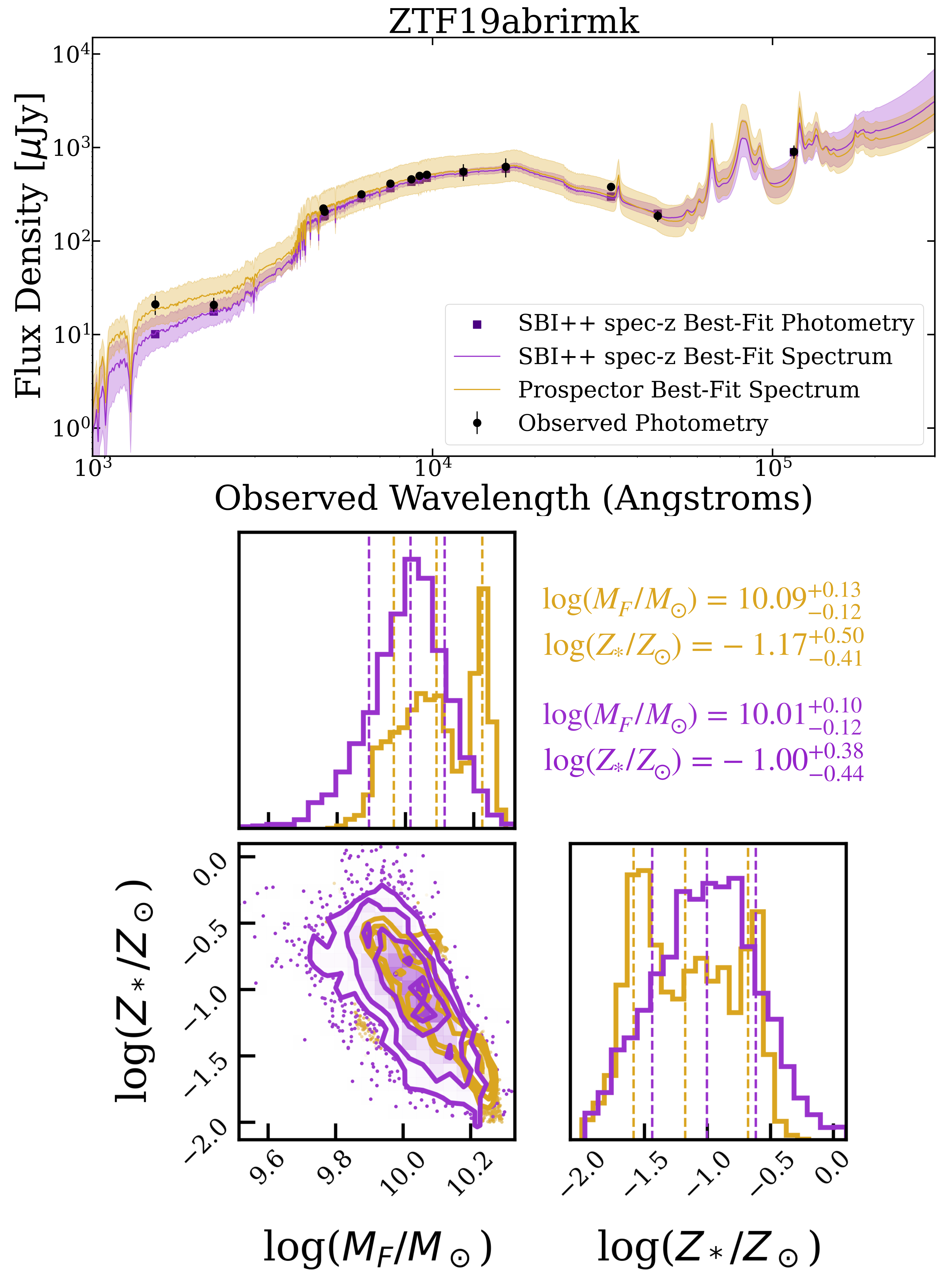}
\includegraphics[width=0.47\textwidth]{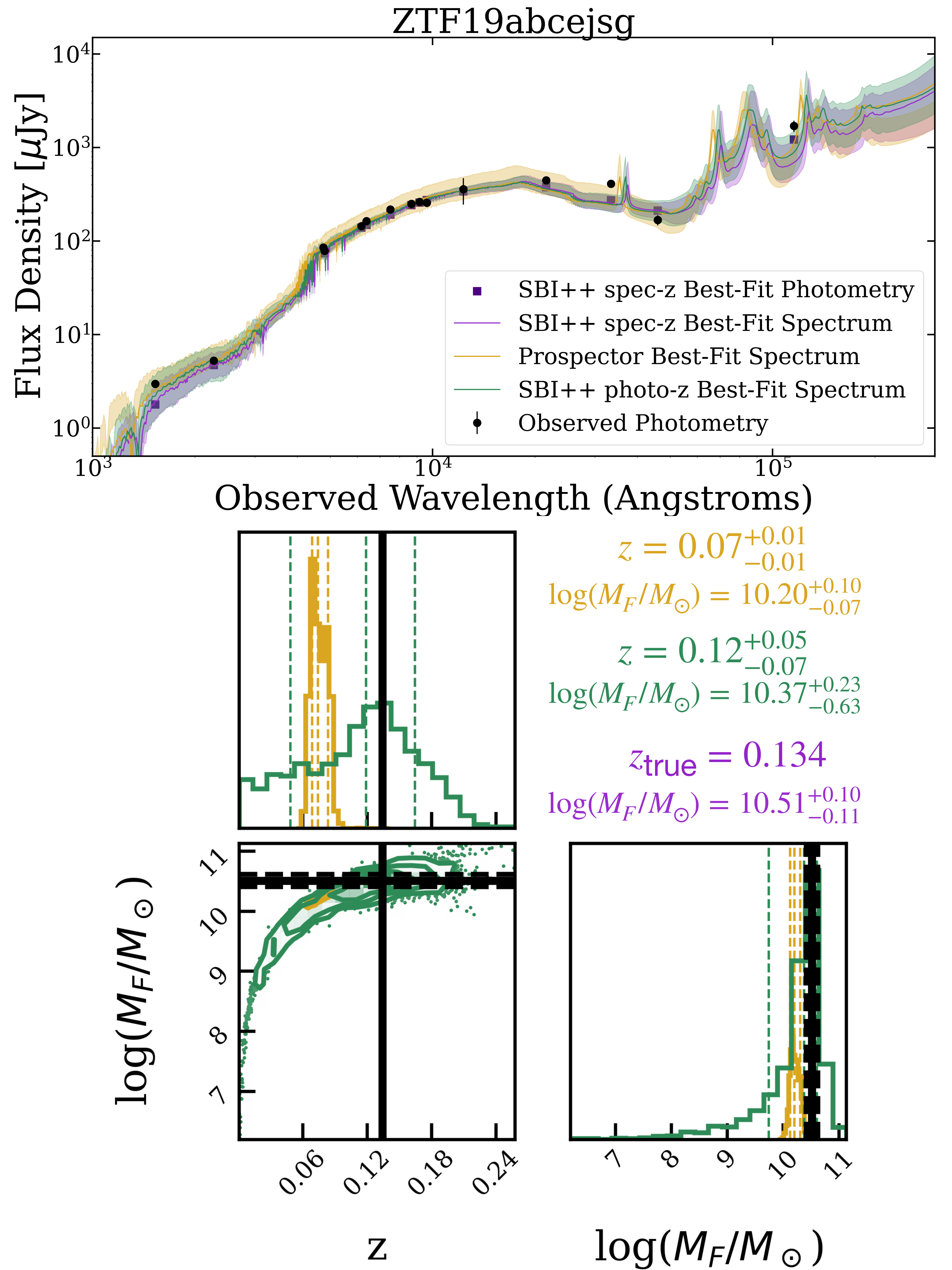}
\caption{ \textit{Left:} A comparison of the model spectral continuum produced from the \texttt{SBI++} spec-$z$ (purple) and \texttt{Prospector} nested sampling fixed-$z$ (yellow) fits to the observed photometry of the host of ZTF19abrirmk (top). When comparing the observed data to the model SEDs with a $\chi^2$ test, we find $\chi^2 = 2\times10^{-5}$ for the \texttt{Prospector} fixed-$z$ model and $\chi^2 = 10^{-4}$ for \texttt{SBI++}. The bottom plot shows the posteriors from the \texttt{Prospector} fit (yellow) and \texttt{SBI++} fit (purple) on $M_F$ and $Z_*$.  Both \texttt{Prospector} and \texttt{SBI++} have consistent posteriors. \textit{Right:} A comparison of the model spectral continuum produced from the \texttt{SBI++} spec-$z$ (purple), \texttt{SBI++} photo-$z$ (green), and \texttt{Prospector} nested sampling (yellow) fits and observed photometry (black circles) for the host of ZTF19abcejsg. When comparing the observed and model SEDs with a $\chi^2$ test, we find $\chi^2 = 3\times10^{-5}$ for the \texttt{Prospector} redshift-free fit and $\chi^2 = 7\times10^{-5}$ for \texttt{SBI++} photo-$z$ fit. In the bottom plot, we show the \texttt{SBI++} photo-$z$ (green) and \texttt{Prospector} nested sampling (yellow) produced posteriors on redshift and $M_F$. We show the 64\% confidence interval (black dashed lines) and median (black line) from the \texttt{SBI++} spec-$z$ fit. While the \texttt{SBI++} photo-$z$ fit has larger uncertainties than the  \texttt{SBI++} spec-$z$ fit, it correctly captures the true redshift and stellar mass of the host. The \texttt{Prospector} nested sampling posteriors are more constrained, but inconsistent with the \texttt{SBI++} spec-$z$ results.}
\label{fig:compare_sed}
\end{figure*}

\subsection{Fitting Observed Data}
\label{sec:run_sbi}
We employ several different methods for fitting the observed photometry of our host population with either the \texttt{SBI++} spec-$z$ or photo-$z$ model, depending on the whether the host has any missing or noisy data. Here, we define ``noisy" data as any photometry with $(\sigma_o - \sigma_m) / \sigma_m > 10$, where $\sigma_o$ is the observed photometric uncertainty and $\sigma_m$ is the photometric uncertainty in the training set at the same magnitude as the observed host in a specified filter. For hosts with complete datasets and no noisy photometry, we fit their observed data with the baseline SBI model. When there is noisy photometry, we produce simulated photometry of the noisy data using a Gaussian distribution, with the mean as the observed point and the standard deviation as the uncertainty. To only allow reasonable simulated photometric measurements, we truncate the Gaussian to be within the range of similar SEDs in the training set, determined through a nearest neighbor search: all training set SEDs with a reduced $\chi^2 \leq 5$. If there are less than 10 SEDs that fit this criteria, we increase the $\chi^2$ threshold by 5 until we reach 10 nearest neighbors. To ensure that we are not including any training set SEDs that result in an extremely high $\chi^2$, we cap the number of nearest neighbors to 200 and the maximum $\chi^2$ to 5000. We find that the median number of nearest neighbors for our host population is 16, with a median $\chi^2 = 20$, validating that the majority of selected nearest neighbor SEDs are reasonable. Then, we fit each combination of the photometry and the simulated photometry of the noisy band(s) with baseline SBI and average over the noisy posterior samples for a final estimate.

For cases when there is missing photometry, we simply do the same nearest neighbor SED search as just described to determine fiducial photometry in the missing bands. We then build a Kernel Density Estimator (KDE) from these SEDs, which we weight by the inverse of their Euclidean distances to the host's observed photometry. Similar to our methods for handling noisy data, we draw samples from the KDE and fit each sample and the observed data with baseline SBI, averaging over the posteriors. Finally, in the case where there is both missing and noisy data, we use a combination of the two previously described methods. For these final three cases (noisy data, missing data, and noisy and missing data), we follow the recommendation from \citet{wlv+2023} and draw 50 samples of each noisy and/or missing band and produce 50 posterior samples per draw (totally 2500 posterior draws). We similarly pull 2500 posterior draws for the fits that do not have missing or noisy data, for uniformity in our sampling. We refer the reader to \citet{wlv+2023} for more details on these methods and the \texttt{SBI++} infrastructure.

\begin{figure}[t]
\centering
\includegraphics[width=0.47\textwidth]{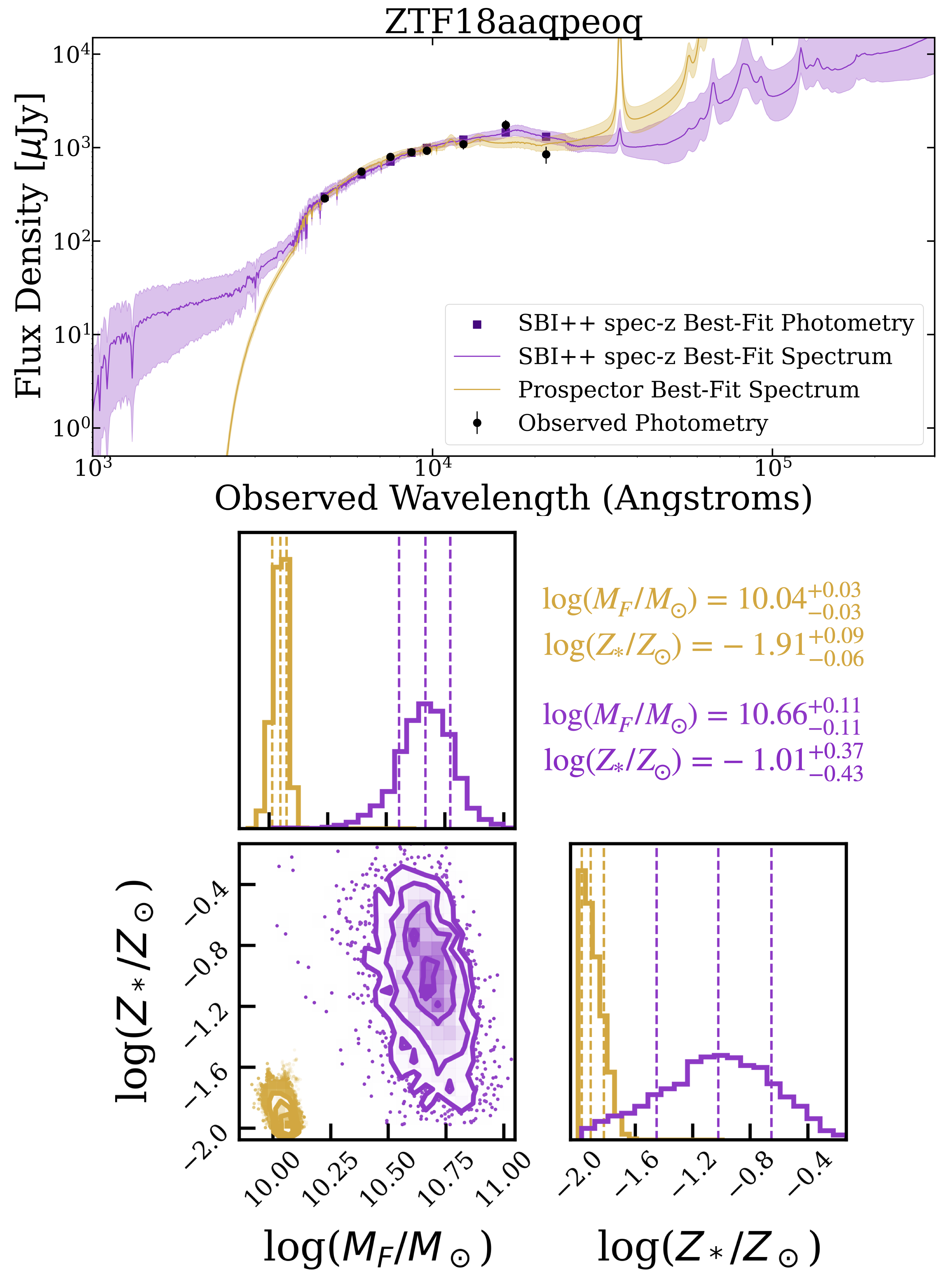}
\vspace{-0.1in}
\caption{The same as Figure \ref{fig:compare_sed} (left panel). The \texttt{Prospector} nested sampling fixed-$z$ and \texttt{SBI++} spec-$z$ fits yielded inconsistent posteriors on stellar population properties for this host: the \texttt{Prospector} nested sampling-inferred $Z_*$ is much lower than expected, given the $M_F$-$Z_*$ relation \citep{gcb+05}. Interestingly, while both \texttt{Prospector} and  \texttt{SBI++} fit the observed data well (both models have a reduced $\chi^2 = 0.0001$), \texttt{Prospector} grossly underestimates the UV flux in comparison to the \texttt{SBI++} model SED, which likely settled on a more neutral UV flux range in comparison to the training set. It appears that  \texttt{Prospector} nested sampling fits with divergent stellar population property estimates from \texttt{SBI++} all underestimate the UV flux, hinting at a failure in the \texttt{Prospector} nested sampling methods to capture the true uncertainty on the model SED.}
\label{fig:bad_prosp}
\end{figure}

As discussed in Section \ref{sec:training}, we employ the spec-$z$ \texttt{SBI++} model when a host has a spectroscopic redshift or a well constrained photometric redshift. For our spectroscopically classified SN sample, we use the best-reported SN redshifts reported on TNS, which are almost always based on host galaxy spectral lines (and thus represent the host redshift), and we fit all hosts with the spec-$z$ model. For the photometrically-classified SN sample, we pull redshift estimates from GLADE, the \textit{Dark Energy Spectroscopic Instrument} (DESI; \citealt{DESI}) survey, DECaLS \citep{Zhou2021}, and Pan-STARRS \citep{PanSTARRS}. All redshifts from GLADE and DESI are spectroscopic. Given the proven accuracy of the DECaLS photometric redshift estimator \citep{Zhou2021}\footnote{\citet{Zhou2021} reports a mean deviation of the photometric redshift to the spectroscopic redshift of 0.02 for galaxies with $z_\textrm{mag} < 21$~mag and $0.05-0.1$ for $z_\textrm{mag} > 21$~mag. We note that 98\% of spectroscopic SN hosts and 90\% of our photometric SN hosts have $r<21$~mag, thus majority of photometric redshifts we have collected from DECaLS should be sufficient for use in our spec-$z$ model.}, we assume that their redshifts are good enough for use in our spec-$z$ model. We assume that any Pan-STARRS redshifts with uncertainty $<0.02$ are suitable to use in our spec-$z$ model. In all other cases, we use our photo-$z$ model. We note that only $\approx12\%$ of photometrically-classified SN hosts are fit with the photo-$z$ model.

In total, we are able to get \texttt{SBI++} fits, and thus stellar population properties, for 9262 hosts. We find that the fits failed for the other 1891 hosts with photometry due to the hosts not having optical photometry (preventing a fit altogether) or having sparse and noisy photometry.  The breakdown of spectroscopically versus photometrically classified transients, and their SN classification, with SED fits is shown in Figure~\ref{fig:sample}. In the following subsection, we discuss accuracy of our SED-fitting techniques as well as comparisons to \texttt{Prospector} SED fits with a Bayesian approach. We point the reader to Sections \ref{sec:specvphot} and \ref{sec:transientcompare} for an overview of our stellar population modeling results.

Finally, we compute some commonly-used physical properties of galaxies from the fitted properties (see Table~\ref{tab:prosp_priors}). For example, using $M_F$ and the SFR ratios, we derive an SFH, an SFR integrated over the most recent 100 Myr (nominally, the present-day SFR; SFR$_{0-100\textrm{Myr}}$), and a mass-weighted age ($t_m$). We furthermore convert $M_F$ to a stellar mass ($M_*$) using the SFH, IMF, and $Z_*$ of each host. We calculate a $V$-band dust extinction magnitude ($A_V$) by multiplying the total optical depth ($\tau_{V,1} + \tau_{V,2}$) by 1.086. Hereafter, we only make stellar population property comparisons between these derived parameters, $z$, and $Z_*$. 

\subsection{Test Cases}
\label{sec:sbi_tests}
To test the accuracy of our \texttt{SBI++} spec-$z$ and photo-$z$ models, we compare the outputs of 100 randomly selected spec-$z$ fits within our spectroscopically classified SN host sample to those derived from traditional \texttt{Prospector} fits with a Bayesian approach and \texttt{SBI++} photo-$z$ fits. We first compare the \texttt{SBI++} spec-$z$ fits to traditional \texttt{Prospector} fits in which we fix the redshift to that of the transient. For our \texttt{Prospector} fits, we use the same model as described in Section \ref{sec:training} and prior distributions as listed in Table \ref{tab:prosp_priors}. We apply a nested sampling fitting routine through \texttt{dynesty} \citep{Dynesty} to sample the prior space and produce posterior distributions of the stellar population properties of interest\footnote{We use 20 CPUs on the Harvard FASRC Cannon cluster for each \texttt{Prospector} fit and find that the majority of fits take $\approx$2~days to finish.}. 

We show an example of the \texttt{Prospector} nested sampling and \texttt{SBI++} spec-$z$ model fits in Figure \ref{fig:compare_sed}. When comparing the inferred stellar population properties between the \texttt{Prospector} nested sampling and \texttt{SBI++} spec-$z$ models, it appears that the majority of fits are in agreement with each other, as shown in Figure~\ref{fig:zfix_p_s}. However, we find that, for a handful of cases, the $Z_*$, $t_m$, and $A_V$ values do substantially differ. In Figure~\ref{fig:bad_prosp}, we highlight an example of a host where the fits resulted in discrepant stellar population properties. When exploring the cause of these differences, we find that these \texttt{Prospector} nested sampling-inferred $Z_*$ fall well-off the $M_F$-$Z_*$ relation enforced in both models, typically landing on either the minimum or maximum of the prior distribution (Table~\ref{tab:prosp_priors}). In contrast, all of the \texttt{SBI++}-inferred metallicities clearly follow the $M_F$-$Z_*$ relation (Figure~\ref{fig:zfix_p_s}). This is the expected behavior, as metallicity is difficult to constrain from photometry alone \citep{jlc+2021}, and, thus, the prior should be returned (as it does for \texttt{SBI++}). It further appears that, in some cases, \texttt{Prospector} nested sampling fits underestimate the UV flux, typically dominated by young, massive star populations, compared to the \texttt{SBI++} model. While no SED-fitting code is designed to predict the SED outside of the observed range (they only know the likelihood), this failure observed in the \texttt{Prospector} fits is likely caused by the metallicity being poorly constrained from the photometry, and the nested sampling fitting routine landing on a local minima in the posterior distribution rather than encapsulating the full posterior uncertainty, as postulated in \citet{wlv+2023}. Indeed, \citet{wang2025} corroborates that locating the global maximum in a likelihood surface in \texttt{Prospector} nested sampling fitting can be difficult. We note that given the well-known age-metallicity-dust degeneracy \citep{Conroy2013}, it is not surprising that the $t_m$ and $A_V$ values inferred from the same galaxies with poor $Z_*$ constraints from \texttt{Prospector} also differ.
\begin{figure*}
\centering
\includegraphics[width=1.0\textwidth]{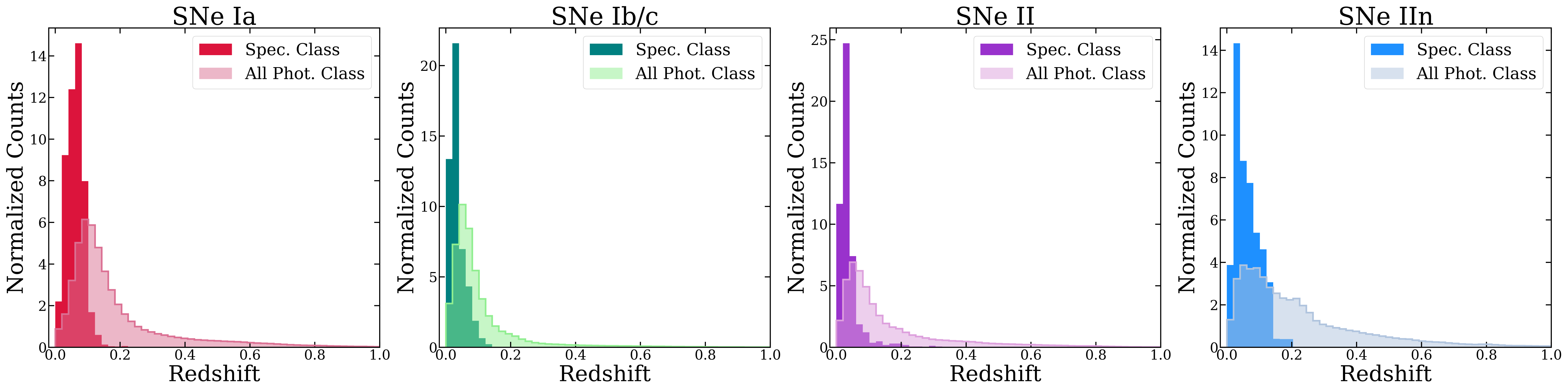}
\vspace{-0.1in}
\caption{The redshift probability distributions (PDFs) of the spectroscopically (darker colors) and photometrically-classified (lighter colors) SNe studied in this work. Within the photometric sample, a handful of hosts were fit with the photo-$z$ \texttt{SBI++} model (see Section \ref{sec:run_sbi}), thus, all redshift posteriors from these fits are included in the PDFs. For the rest of photometric SN sample, we include 2500 realizations determined through random sampling from a Normal distribution, using the redshift mean and standard deviation of the host provided by DECaLS or GLADE. For all SNe classes, the photometric sample traces higher redshifts than the spectroscopic sample.}
\label{fig:redshift}
\end{figure*}

We next compare \texttt{Prospector} nested sampling fits with redshift as a free parameter to the \texttt{SBI++} photo-$z$ fits, using the results from the \texttt{SBI++} spec-$z$ fits as the true inferred properties. In Figure \ref{fig:compare_sed}, we display an example of the \texttt{SBI++} spec-$z$ and photo-$z$ fits. We show the comparison between the \texttt{SBI++} photo-$z$, spec-$z$, and \texttt{Prospector} nested sampling samples in Figure~\ref{fig:zfree_p_s}. The most notable difference between the \texttt{Prospector}  nested sampling and \texttt{SBI++} photo-$z$ fits is in the redshift estimation. As also shown in \citet{wlv+2023}, \texttt{SBI++} tends to produce more realistic uncertainties on redshift than \texttt{Prospector} nested sampling, which, occasionally, gives too stringent constraints on an incorrect redshift estimate. In turn, we find that many redshift-dependent properties, such as stellar mass and SFR, are also incorrect in the \texttt{Prospector} nested sampling fits. The \texttt{SBI++} photo-$z$ fits appear to have wider uncertainties on these properties, which typically encapsulate the true estimate. Finally, we note that there are a handful of \texttt{Prospector} nested sampling fits that fall well off the $M_F$-$Z_*$ relation, which also causes discrepancies in $t_m$ and $A_V$ for the same reasons as was previously discussed for the fixed redshift \texttt{Prospector} fits.

Overall, we find that our \texttt{SBI++} fits have more realistic posteriors on stellar population properties and give better calibrated estimates (see Appendix \ref{app:prosp_sbi}, Figure \ref{fig:coverage}) on the posterior than \texttt{Prospector} nested sampling, thus validating our use of this novel technique in \texttt{FrankenBlast}. 

\section{Spectroscopic vs. Photometric Classes}
\label{sec:specvphot}
In this section, we compare the host stellar population properties between spectroscopically and photometrically classified transients within the same transient class. Our primary goal is to understand if host stellar population properties are useful in determining how ``pure" our photometric samples are, which will be relevant for generating large and complete transient samples in the Rubin era. For example, we may expect that a photometric sample with a large fraction of contamination from misclassified events will have stellar population property distributions that deviate significantly from the spectroscopic sample. On the other hand, if the photometric samples lie at different redshifts than the spectroscopic samples, this may also affect the shape of their distributions, as several galaxy properties are redshift-dependent. Thus, we also seek to explore the affect of intrinsic biases within our samples that may influence their observed host properties.

We note that we only focus on the host populations of SNe~Ia, Ib/c, II, and IIn. We exclude SLSNe-I in this part of the analysis. In addition to being a small fraction of our sample, SLSNe-I hosts are generally quite faint; \citet{schulze2021} estimates an average $r$-band magnitude of $\approx23$~mag, which goes beyond the limiting magnitude of the optical surveys used in this study. Thus, its possible that \texttt{Pröst} did not find the correct host for some of this sample, given that a number of SLSNe-I hosts were likely undetected. Indeed, the median ``missed catalog probability" for the spectroscopic SLSNe-I sample is $0.13^{+0.08}_{-0.09}$ (median and 68\% confidence interval), compared to $0.007^{+0.13}_{-0.007}$  for the rest of the host associations.

We detail comparisons of redshift, stellar mass, star and formation, and dust extinction in the following subsections. While we also infer stellar population age and stellar metallicity for these populations, we do not think these properties are uniquely useful in determining whether a photometric SN population is contaminated. For example, the age distribution will likely closely follow that of the redshift distribution. Additionally, the \citet{gcb+05} mass-metallicity prior used in the \texttt{SBI++} training set closely links $Z_*$ to $M_*$ (as also seen in our results; Figures \ref{fig:zfix_p_s}-\ref{fig:zfree_p_s}). We show all population medians and 68\% confidence intervals on all SN populations (divided by spectroscopic versus photometric samples) in Table \ref{tab:sp_results}.

\begin{figure*}
\centering
\includegraphics[width=1.0\textwidth]{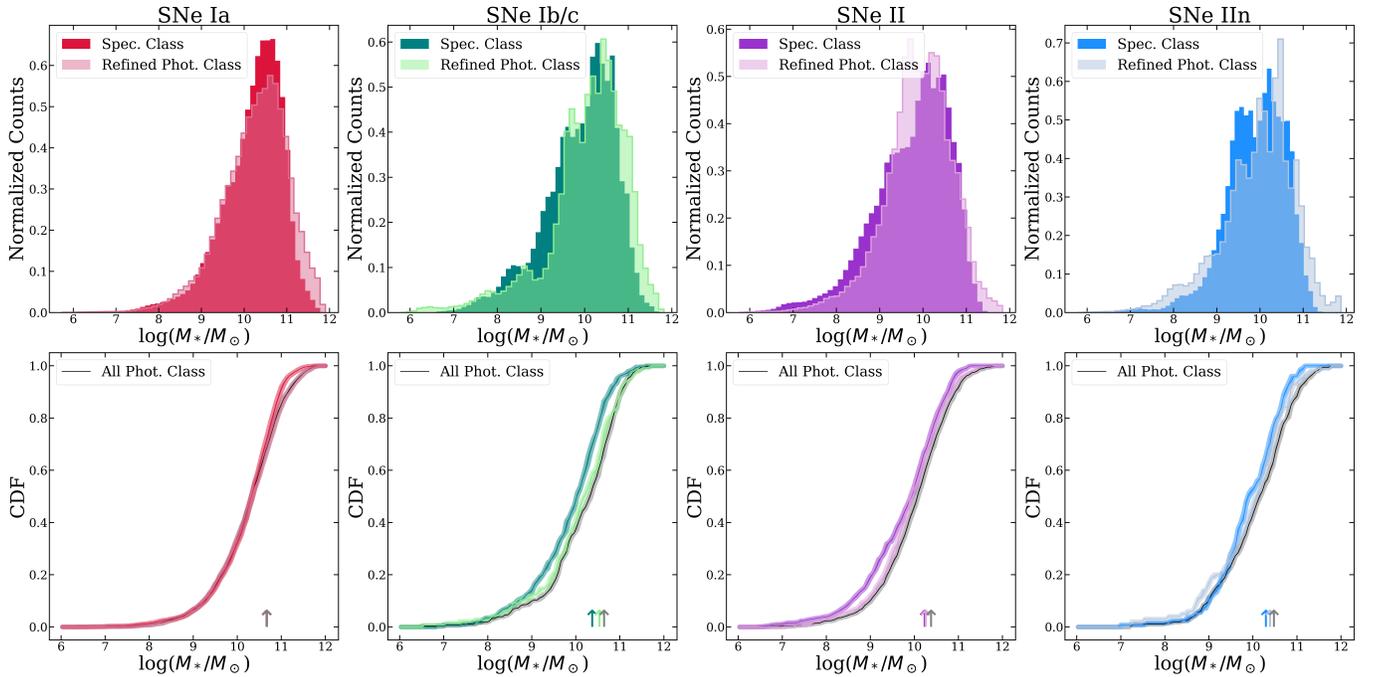}
\vspace{-0.1in}
\caption{\textit{Top row:} The PDF in stellar mass for the spectroscopic SNe samples (darker colors) and refined photometric SNe samples (lighter colors). \textit{Bottom row:} The stellar mass CDFs for the aforementioned samples, as well as the full photometric SNe samples (dark grey). The arrows point to the medians of the distributions. We generally find that the photometric SNe hosts lie at higher stellar mass than the spectroscopic SNe hosts, which is likely caused by a combination of contamination of other transient events, incorrect host associations, and non-detectability of low-mass hosts at higher redshift.}
\label{fig:mass_s_p}
\end{figure*}

\subsection{Redshift}
\label{sec:redshift}
We begin by analyzing the redshift distributions of our spectroscopic and photometric SN samples and showcase these distributions in Figure \ref{fig:redshift}. To build the distributions, we include all 2500 posterior draws of redshift from the \texttt{SBI++} fit. For spectroscopic SN hosts, this is simply drawing the same redshift used in the fit 2500 times. For photometric SN hosts fit with the spec-\textit{z} model, we instead randomly sample 2500 redshifts from a Normal distribution, using the mean redshift and standard deviation provided by the survey where we obtained the redshift estimate. We list the redshift population medians and 68\% confidence intervals for each SN class, divided in spectroscopic and photometric class designations, in Table \ref{tab:sp_results}. For all SN classes, we find that the spectroscopic sample lies at lower redshifts than the photometric sample. This is unsurprising, given that lower redshift transients tend to be brighter and, thus, more likely to be followed up spectroscopically. While the differences in redshifts of the spectroscopic and photometric samples are not useful in identifying the fraction of potentially misclassified events, this will have an effect on distributions of other stellar population properties that are important in distinguishing potential misclassifications. We discuss the impact of redshift in more detail in Sections \ref{sec:mass}-\ref{sec:sf}.

\subsection{Stellar Mass}
\label{sec:mass}
We next compare the stellar masses of the spectroscopic and photometric SN samples and show the posterior and cumulative distribution functions (CDFs) in Figure \ref{fig:mass_s_p}. As was done with redshift, we draw 2500 posterior samples of stellar mass inferred from each \texttt{SBI++} fit. To encapsulate uncertainty on the CDFs, we build 2500 distributions and determine the median CDF and 68\% confidence interval on the CDF from the realizations. For all SN populations, we find that the photometric sample leans towards higher stellar mass hosts than the respective spectroscopic sample, with the effect appearing more dramatic for the CCSNe populations (SNe Ib/c, II, and IIn) than the SNe~Ia. We confirm the sample differences quantitatively with and Anderson-Darling (AD) rejection test, with the null hypothesis that the spectroscopic and photometric stellar mass distributions are derived from the same underlying distribution. We perform an AD test for each of the 2500 realizations on the CDF between spectroscopic and photometric classes to build a distribution of probabilities ($P_{AD}$). If the majority of $P_{AD} < 0.05$, we reject the null hypothesis. For all SN classes, we find that 100\% of $P_{AD}<0.05$, thereby rejecting the null hypothesis.

Initially, we believe that the differences found between the spectroscopic and photometric CCSNe samples might be caused by a combination of SNe~Ia contaminating the CCSNe photometric samples and incorrect host associations in the photometric sample. Indeed, SNe~Ia tend to occur in galaxies with higher stellar mass than CCSNe populations (\citealt{qz2024}; and Section \ref{sec:transientcompare}). Moreover, given the lower stellar masses and faintness of CCSNe hosts \citep{schulze2021}, it is possible that a number of true CCSNe hosts are undetected in the wide-field surveys used in this study. In these cases, \texttt{Pröst} may have selected a potentially higher mass, brighter galaxy as the host of the event. Thus, we restrict CCSNe photometric populations (and SNe Ia, for posterity) to only those with high classification and host association probabilities as we will describe below. 

To determine the events with the best photometric classifications, we acquire the \texttt{Superphot+} classification probabilities (described in Section \ref{sec:sample}). We note that these are not true probabilities, but rather ``pseudo-probabilities", as they are not calibrated \citep{deSoto2024}. \citet{deSoto2024} does, however, explore how well-calibrated the pseudo-probabilities are by determining if the classification probability matches the fraction of events that are correctly classified at that probability. They find that SN~Ib/c pseudo-probabilities tend to be overconfident, SNe~Ia pseudo-probabilities are under-confident, and the other SN classes do not have bias. We decide to clip all photometric SNe host samples to those with classification pseudo-probabilities $\geq50\%$, as this is the point at which the majority of events within our photometric samples should be correctly classified, without restricting the samples by so much that they become too small for statistical analysis. We emphasize that this cut will lead to varying levels of contamination in each SN class: e.g., SNe II will have less contamination than SNe Ib/c. 

For the host association cut, we formulate a new probability ($P_\textrm{chosen}$) that represents the posterior odds that the chosen host is the true host versus the host not being in the catalog at all:
\begin{equation}
    P_\textrm{chosen} = P_\textrm{Prost} / (1 - P_\textrm{any}),
\end{equation}
where $P_\textrm{Prost}$ is the \texttt{Pröst} host probability, and $P_\textrm{any} = 1 - (P_\textrm{faint} + P_\textrm{outside})$. Here, $P_\textrm{faint}$ is the probability of an unobserved galaxy being the true host, $P_\textrm{outside}$ is the probability of the true host being outside the search radius, and $P_\textrm{any}$ is the probability that one of the galaxies within the search radius and detected in the catalog is the true host. We cut out any hosts where $P_\textrm{chosen}/P_\textrm{Prost} < 5$, or where $P_\textrm{any} <80\%$. In total, these cuts reduce the photometric SNe Ia sample to 2527 events, the photometric SNe Ib/c sample to 143 events, the photometric SNe II sample to 410 events, and the photometric SNe IIn sample to 100 events. From here on, we refer to the reduced photometric samples as the ``refined" photometric samples. The stellar mass PDF and CDFs of the refined samples are also shown in Figure \ref{fig:mass_s_p}. We note that $\approx 10\%$ of the refined photometric sample is fit with the \texttt{SBI++} photo-$z$ model.

For all CCSNe types, we find that host stellar mass distributions of the refined photometric sample shift closer to the spectroscopic sample than that full photometric sample, but, visually, still lie at slightly higher stellar masses than the spectroscopic sample. We confirm the statistical deviation for SNe~Ia, SNe Ib/c, and SNe II with AD tests, where we are able to reject the null hypothesis that the hosts of photometrically-classified follow the same stellar mass distribution as the spectroscopic sample (100\% of $P_{AD}<0.05$ for SNe Ia, 97\% for SNe Ib/c, and 99\% for SNe II). For SNe IIn, we determine that only 12\% of AD tests result in $P_{AD}<0.05$; thus, we are unable to reject the null hypothesis.  

\begin{figure*}
\centering
\includegraphics[width=1.0\textwidth]{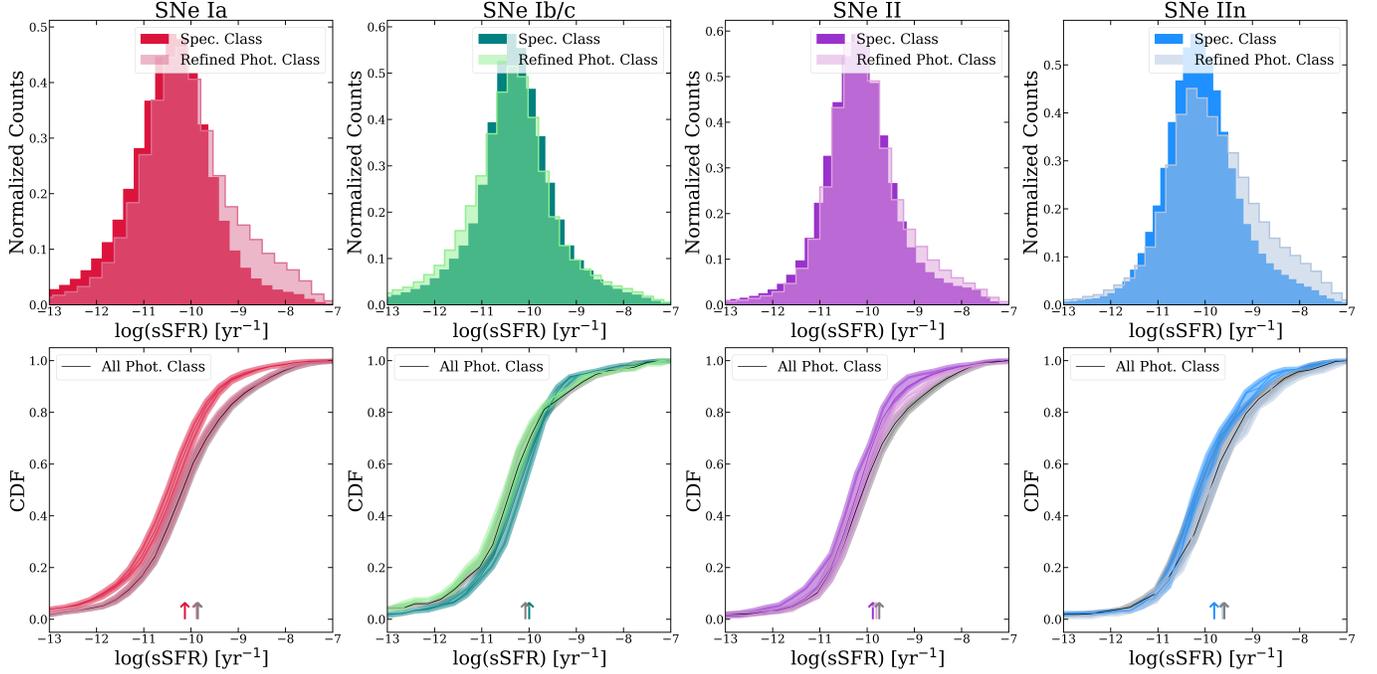}
\vspace{-0.1in}
\caption{The same as Figure \ref{fig:mass_s_p}, except for sSFR. We find that for all SNe, except SNe Ib/c, the full photometric sample and refined photometric samples lean towards higher sSFRs.}
\label{fig:ssfr_s_p}
\end{figure*}

Given that there should be less contamination within the refined photometric sample (although, there still might be considerable contamination in the SNe Ib/c sample, which we explore in Section \ref{sec:dust}), we believe that this may partially be due to lack of detectability of low-mass hosts at higher redshift. Although we may expect that that populations of galaxy stellar mass decreases with increasing redshift \citep{erb2006, speagle2014, vanderWel2014, whitaker2014, Leja2019, leja2022}, the observational bias in our observed host population may make it more difficult to observe these faint, low mass hosts. As a test of this hypothesis, we produce 2 million mock \texttt{Prospector} SEDs (using the same model as discussed in Section \ref{sec:training}) at $z<0.2$ (where the majority of the photometric SNe lie) and $M_*=10^8 M_\odot$. Indeed, we find that $>50\%$ of mock galaxies have an $r$-band magnitude $>23.5$ mag, which is the limiting magnitude of Pan-STARRS. 

We further contemplate whether the distributions could be disparate due to host-based selection effects. As shown in \citet{fsn+2017}, the discovery of transients becomes more difficult if their hosts have higher surface brightness. This may mean that spectroscopic classification is more challenging  (or less desirable) for transients centrally located in bright, possibly higher stellar mass galaxies, and, thus, spectroscopic samples may be naturally biased towards lower stellar mass galaxies than photometric samples. We note that the majority (60\%) of our spectroscopic SN sample comes from BTS, which is fairly complete for transients down to 19~mag \citep{perley2020}, so we do not expect that this portion of our spectroscopic sample is biased by host galaxy surface brightness or host stellar mass. To test whether the other 40\% of the spectroscopic sample is influenced by host galaxy properties, we compare the redshifts, host $r$-band magnitudes, and stellar masses between the BTS and non-BTS spectroscopic samples. We find that all properties are very similar across all SN classes, except for the SNe Ia stellar mass distributions where we find that the non-BTS sample leans towards slightly lower stellar masses than the BTS sample (although, the difference is small). We confirm this with an AD test. This may suggest that our non-BTS spectroscopic SNe Ia  sample is biased against transients in higher stellar mass galaxies, which could partially explain why our spectroscopic SNe Ia hosts trend towards lower stellar masses than our photometric SNe Ia host sample. However, this effect will be small in comparison to the redshift effect between the spectroscopic and photometric samples we explored previously. Given the lack of evidence that our BTS and non-BTS CCSNe host samples are distinct, we do not think that host-based selection effects are impacting the differences between our spectroscopic and photometric CCSNe host populations at all. We do strongly caution that fully understanding the impact of host-based selection effects on spectroscopic samples requires a much deeper analysis on SN rates than what we can perform here and, thus, this argument is mainly speculative.

Finally, we determine that we do not require many more low-mass hosts in the SNe Ib/c and II samples for their stellar mass distributions to become more statistically similar. With the addition of only seven more $10^8 M_\odot$ host in the SNe Ib/c refined  photometric sample and ten more $10^8 M_\odot$ hosts in the SNe II refined photometric sample, we find that the fraction of $P_{AD}<0.05$ drops below 50\%. Since this is such a small fraction of the refined SNe Ib/c and II photometric samples ($<5\%$), we bootstrap to test whether the initial AD tests between the refined photometric and spectroscopic samples are swayed by small number statistics. Within the respective spectroscopic SN samples, we randomly sample (with replacement) the number of SNe in the refined photometric samples 1000 times, pulling all 2500 stellar mass posteriors for the sampled hosts. We then perform 2500 AD tests and calculate the fraction of AD tests with $P_\textrm{AD} < 0.05$  for each of the 1000 samples. For both SNe Ib/c and II, the majority of tests can still be rejected (653/1000 for Ib/c and 774/1000 for II). This either implies that our initial AD results are not influenced by the smaller sample sizes or that the AD test is just too sensitive to the tails of the distributions, leading to the test rejections.

\subsection{Star Formation}
\label{sec:sf}
We now explore possible differences in star formation between the spectroscopic and photometric SN hosts. To fairly compare the amount of active star formation in the hosts, we create distributions of present-day specific SFR (sSFR = SFR$_{0-100\textrm{Myr}}$/$M_*$), which is advantageous to use when analyzing galaxies over a large stellar mass range \citep{Leja2019, leja2022}. We produce posterior and cumulative distributions in the same way as for stellar mass (Section \ref{sec:mass}) and show these distributions in Figure \ref{fig:ssfr_s_p}. It appears that most of the photometric samples trace higher sSFRs than the spectroscopic samples. When comparing spectroscopic and photometric SN host sSFR distributions with our AD test, we are able to reject the null hypothesis that the photometric SN host sSFR distribution traces the same sSFR distribution as the spectroscopic SN hosts for all SN classes except SNe Ib/c (100\% $P_{AD} < 0.05$ for SNe Ia and SNe II, 10\% for SNe Ib/c, and 86\% for SNe IIn). When we just compare the spectroscopic sample to the refined photometric sample (Section \ref{sec:mass}), we are only able to reject the null hypothesis for SNe Ia and SNe II, with less significance than the result of the AD test with the full photometric sample for SNe II (67\% of $P_{AD} < 0.05$). SNe IIn no longer have statistical evidence for differing sSFR distributions between the spectroscopic and refined photometric samples, as only 40\% of $P_{AD} < 0.05$.

We assume that the photometric SN hosts will have higher sSFRs than the spectroscopic SN hosts, given that star formation increases with increasing redshift \citep{speagle2014, whitaker2014, leja2022}. To further explore this concept and test whether redshift can solely account for the differences observed in the SN Ia and II samples, we compare sSFRs within a simulated field galaxy population. To build this population, we utilize the \citet{leja2022} SFR-$M_*$ field galaxy probability density function and the methods outlined in \citet{hm2025} to randomly generate 16,000 stellar masses and SFRs at each of the redshifts within the spectroscopic and refined photometric SN Ia and II samples\footnote{We use the median redshift in the cases where redshift was determined in the \texttt{SBI++} fit.}. In order to be consistent with the field galaxy population observed within the surveys used here, we remove any simulated galaxies with an optical magnitude $>23.5$~mag. We provide further details on the \citet{hm2025} method for generating a simulated galaxy population in Section \ref{sec:multi}.

\begin{figure*}[t]
\centering
\includegraphics[width=1.0\textwidth]{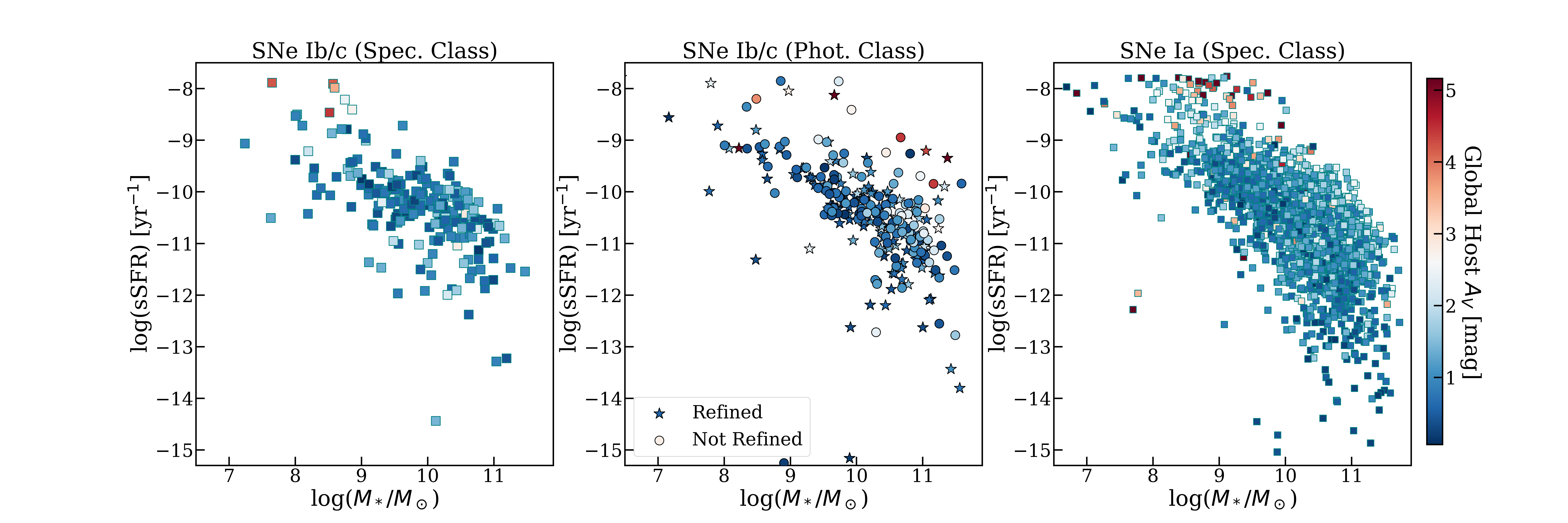}
\vspace{-0.2in}
\caption{The host galaxy sSFRs versus stellar masses, colored by the global host dust extinction ($A_V$), for the spectroscopically-classified SNe Ib/c (left), photometrically-classified SNe Ib/c (middle), and spectroscopically-classified SNe Ia (right). The photometrically-classified SNe are further broken down into whether the SN is in the refined sample (stars) or not (circles). The spectroscopically-classified SNe Ib/c generally have low $A_V$ at higher stellar mass, whereas there are a few cases of photometrically-classified SNe Ib/c with high host $A_V$ at high stellar mass. These properties are more similar to that of SNe~Ia, suggesting that these events may actually be misclassified SNe~Ia, which can appear redder, and thus more similar to SNe Ib/c, at peak luminosity due to dust in the host.}
\label{fig:Ibc_dust}
\end{figure*}

After the simulated galaxy populations are created, we randomly pull 5000 stellar masses and SFRs at each SN redshift and build 5000 realizations in sSFR for each of the samples correlating with the redshifts of the SN spectroscopic and refined photometric samples. We then perform 5000 AD tests, with the null hypothesis that the simulated galaxy population following the redshift distribution of a refined photometric SNe sample has the same underlying sSFR distribution as that of the respective spectroscopic SNe sample. We find that the simulated population at the same redshifts as the refined photometric SNe Ia trends towards higher sSFRs (population median $1.3\times10^{-11}$yr$^{-1}$) than the population congruent with the spectroscopic SNe Ia redshifts (population median $1.1\times10^{-11}$yr$^{-1}$) and we can reject 81\% of AD tests. This strongly suggests that the higher sSFRs observed in the refined photometric SNe Ia sample compared to the spectroscopic sample can be explained solely by their redshift differences. Meanwhile, we find that we are only able to reject 29\% of AD tests between the sSFRs of the field galaxy populations at the same redshifts as the refined photometric and spectroscopic SNe II samples. This naturally brings into question whether there is a bias in the spectroscopic SNe II sample towards less star-forming hosts. Indeed, it is possible that this is an example of a host-based selection effect impacting our spectroscopic sample, where we may assume that it is more difficult to discover or spectroscopically classify bluer transients (like SNe II) in more star-forming galaxies, where the hosts' bluer continua may dominate over the transient emission. However, quantifying this effect is difficult to understand without recovering their rates, which goes beyond the scope of this paper. We do note that the differences in sSFR between the refined photometric and spectroscopic samples are subtle. Thus, this observed difference may be irrelevant in overall progenitor interpretation between the broad classes explored here.

\subsection{Dust Extinction}
\label{sec:dust}
We finally explore host galaxy dust extinction ($A_V$) between our spectroscopic and photometric SN Ib/c population. We decide to solely focus on this SN population as their photometric sample likely has the highest rates of contamination and their main contaminants, SNe Ia, are often misclassified as SNe Ib/c due to redder-than-expected $g$-$r$ colors at peak \citep{deSoto2024}. Because dust reddens SNe colors, we may expect higher rates of misclassifications when the environment has more dust. Additionally, environmental line-of-sight $A_V$ is not corrected in the light curves classified by \citet{deSoto2024}, meriting further exploration to understand if this affects contamination rates.  We do not expect that contaminants in the other SNe population (Ia, II, and IIn) are similarly misclassified due to higher $A_V$ in the host, and therefore we leave them out of this analysis.

In Figure \ref{fig:Ibc_dust}, we highlight the global host $A_V$ in the spectroscopic SNe Ib/c sample, compared alongside sSFR and stellar mass. We find that their hosts generally have minimal dust (population median $A_V\approx0.8$~mag; Table \ref{tab:sp_results}), with a few hosts in lower mass, dwarf galaxies ($<10^9 M_\odot$) with high sSFR having $A_V \gtrsim 2$~mag. This trend is generally expected, as the amount of star formation in a galaxy relates to the amount of dust \citep{one+2017, pxn+2023}. The hosts of both the entire and refined photometric SNe Ib/c sample also appear to follow this trend (Figure \ref{fig:Ibc_dust}). However, there are a handful of photometric SN Ib/c hosts with higher stellar mass ($>10^{10}M_\odot$) and higher sSFR ($\gtrsim 10^{-10}$~yr$^{-1}$) with high $A_V \gtrsim 3$~mag ($\sim 7$~events). Indeed, we find that the population median of the entire photometric SNe Ib/c sample is $A_V\approx0.9$~mag, and skews towards dustier galaxies than the spectroscopic sample. While the $A_V$ medians and distributions are more similar between the refined photometric and spectroscopic SNe Ib/c samples, we still find a few outlier hosts with high stellar mass, high sSFR, and high $A_V$. Interestingly, there are a number of hosts in this stellar mass and sSFR regime in the spectroscopic SN Ia sample with similarly high $A_V$ to these outliers (Figure \ref{fig:Ibc_dust}). When inspecting the spectroscopic SNe II and IIn hosts, we do not find hosts with these same properties. Thus, their host properties alone would suggest that they are more similar to the SN Ia population than the SNe Ib/c population. It further is probable that their host's higher $A_V$ may have severely reddened the SN colors, explaining why these events are misclassified as SNe Ib/c. Indeed, when inspecting their SN properties more closely, we find that several of these events do appear more similar SNe Ia from their SN lightcurves and brightness, while the rest appear to have very red lightcurves, suggesting that they may have been severely reddened from the high host $A_V$. We take this as the strongest evidence for understanding which events are misclassified within our SNe Ib/c photometric sample, as the host properties are so distinct from the spectroscopic Ib/c sample, yet more similar to that of SNe Ia.

\begin{deluxetable*}{l|lllllll}[t]
\tabletypesize{\footnotesize}
\tablecolumns{10}
\tablewidth{0pc}
\tablecaption{Stellar Population Properties
\label{tab:sp_results}}
\tablehead{
\colhead{Class} &
\colhead{Redshift} &
\colhead{$\log(M_*/M_\odot)$} &
\colhead{SFR [$M_\odot$~yr$^{-1}$]} &
\colhead{log(sSFR) [yr$^{-1}$] } &
\colhead{$t_m$ [Gyr]} &
\colhead{$\log(Z_*/Z_\odot$)} &
\colhead{$A_V$ [mag]} 
}
\startdata
SNe Ia (S) & $0.06^{+0.02}_{-0.03}$ & $10.34^{+0.53}_{-0.8}$ & $0.61^{+2.21}_{-0.52}$ & $-10.44^{+0.81}_{-0.99}$ & $7.05^{+1.53}_{-2.55}$ & $-0.69^{+0.49}_{-0.72}$ & $0.96^{+1.09}_{-0.6}$ \\
SNe Ia (P - all) & $0.12^{+0.11}_{-0.04}$ & $10.34^{+0.63}_{-0.83}$ & $1.24^{+7.91}_{-1.0}$ & $-10.16^{+1.1}_{-0.9}$ & $6.33^{+1.61}_{-3.71}$ & $-0.57^{+0.46}_{-0.72}$ & $0.99^{+1.42}_{-0.66}$ \\ 
SNe Ia (P - refined) & $0.12^{+0.1}_{-0.04}$ & $10.35^{+0.64}_{-0.84}$ & $1.19^{+7.15}_{-0.95}$ & $-10.18^{+1.09}_{-0.89}$ & $6.38^{+1.58}_{-3.64}$ & $-0.56^{+0.45}_{-0.72}$ & $0.96^{+1.35}_{-0.64}$ \\ \hline
SNe Ib/c (S) & $0.03^{+0.03}_{-0.01}$ & $10.04^{+0.62}_{-0.94}$ & $0.48^{+1.77}_{-0.42}$ & $-10.31^{+0.78}_{-0.79}$ & $6.8^{+1.48}_{-3.19}$ & $-0.73^{+0.52}_{-0.74}$ & $0.78^{+0.9}_{-0.45}$ \\
SNe Ib/c (P - all) & $0.06^{+0.06}_{-0.02}$ & $10.31^{+0.62}_{-0.88}$ & $0.6^{+2.34}_{-0.49}$ & $-10.39^{+0.87}_{-0.92}$ & $6.86^{+1.47}_{-2.83}$ & $-0.58^{+0.43}_{-0.73}$ & $0.87^{+1.13}_{-0.56}$ \\ 
SNe Ib/c (P - refined) & $0.06^{+0.04}_{-0.02}$ & $10.21^{+0.67}_{-0.84}$ & $0.47^{+1.71}_{-0.39}$ & $-10.38^{+0.83}_{-1.0}$ & $6.95^{+1.51}_{-2.99}$ & $-0.62^{+0.46}_{-0.77}$ & $0.81^{+1.07}_{-0.51}$ \\ \hline
SNe II (S) & $0.03^{+0.02}_{-0.01}$ & $9.9^{+0.68}_{-1.04}$ & $0.43^{+1.59}_{-0.36}$ & $-10.19^{+0.75}_{-0.71}$ & $6.66^{+1.44}_{-3.4}$ & $-0.78^{+0.53}_{-0.74}$ & $0.77^{+0.94}_{-0.48}$ \\
SNe II (P - all) & $0.08^{+0.16}_{-0.04}$ & $10.05^{+0.71}_{-0.79}$ & $0.96^{+6.0}_{-0.77}$ & $-10.05^{+1.05}_{-0.74}$ & $6.24^{+1.62}_{-3.76}$ & $-0.65^{+0.51}_{-0.73}$ & $0.89^{+1.34}_{-0.58}$ \\ 
SNe II (P - refined) & $0.06^{+0.05}_{-0.02}$ & $9.93^{+0.7}_{-0.73}$ & $0.67^{+2.56}_{-0.52}$ & $-10.11^{+0.88}_{-0.65}$ & $6.46^{+1.45}_{-3.38}$ & $-0.71^{+0.49}_{-0.7}$ & $0.77^{+1.01}_{-0.48}$ \\ \hline
SNe IIn (S) & $0.05^{+0.05}_{-0.03}$ & $9.97^{+0.62}_{-0.69}$ & $0.67^{+2.62}_{-0.52}$ & $-10.12^{+0.81}_{-0.7}$ & $6.6^{+1.42}_{-3.39}$ & $-0.86^{+0.6}_{-0.7}$ & $0.97^{+1.09}_{-0.62}$ \\
SNe IIn (P - all) & $0.15^{+0.19}_{-0.09}$ & $10.15^{+0.72}_{-0.87}$ & $1.46^{+12.8}_{-1.21}$ & $-9.9^{+1.07}_{-0.85}$ & $5.69^{+1.85}_{-3.85}$ & $-0.64^{+0.53}_{-0.8}$ & $1.02^{+1.58}_{-0.69}$ \\
SNe IIn (P - refined) & $0.12^{+0.16}_{-0.08}$ & $10.06^{+0.67}_{-0.91}$ & $1.18^{+9.24}_{-1.02}$ & $-9.94^{+1.13}_{-0.84}$ & $5.96^{+1.86}_{-3.94}$ & $-0.66^{+0.52}_{-0.77}$ & $0.96^{+1.48}_{-0.61}$ \\ 
\enddata
\tablecomments{The population medians and 68\% confidence interval on redshift, stellar mass ($M_*)$, SFR, sSFR, mass-weighted age ($t_m$), stellar metallicity ($Z_*$), and dust extinction ($A_V$) for the spectroscopically-classied SNe (S), the entire set of photometrically-classified SNe (P - all), and the refined photometric sample (P - refined; Section \ref{sec:mass}).}
\end{deluxetable*}

\begin{figure*}[t]
\centering
\includegraphics[width=0.9\textwidth]{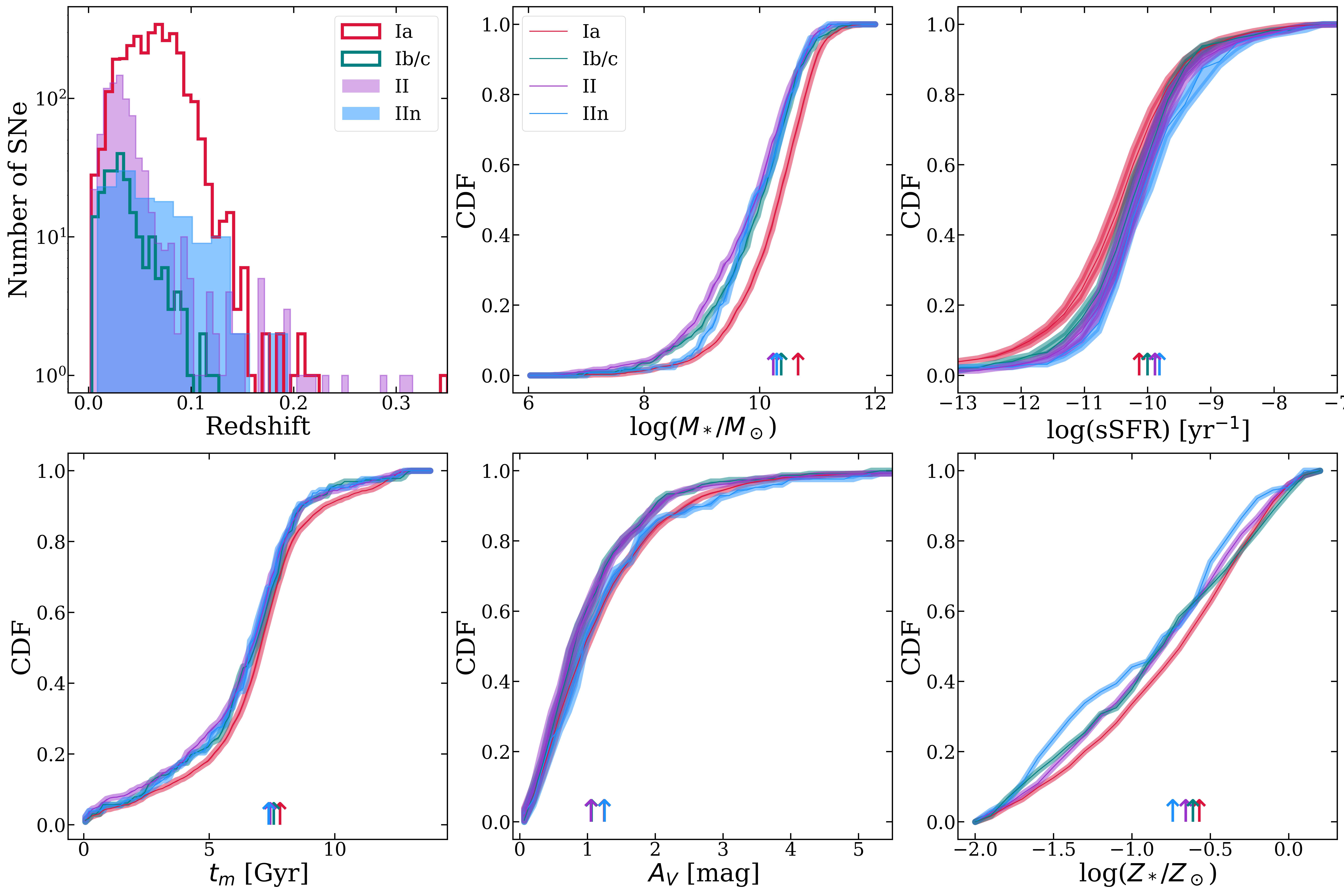}
\vspace{-0.1in}
\caption{\textit{From left to right top to bottom:} The redshift distributions of spectroscopically-classified SNe Ia (red), Ib/c (green), II (purple), and IIn (blue), as well as cumulative distributions (CDFs) in stellar mass ($M_*$), sSFR, mass-weighted age ($t_m$), dust extinction ($A_V$), and stellar metallicity ($Z_*$). The arrows indicate the median in the CDFs.}
\label{fig:compare}
\end{figure*}

\section{Comparisons between Transient Classes}
\label{sec:transientcompare}
In this section, we compare the host properties between different SN classes to better understand if they prefer distinct environments. This is not only useful for progenitor inference (i.e., understanding their delay time distributions and formation channels), but also for discerning trends that may help to better classify the multitude of SNe that will be discovered by, e.g., Rubin and \textit{Roman}. We only focus on the environments of our spectroscopic SN sample for the most accurate representation of their SN class properties. In Section \ref{sec:indiv}, we compare distributions of individual stellar population properties between SNe Ia, Ib/c, II, and IIn hosts. We explore hosts in a multivariate SFR-$M_*$ plane in Section \ref{sec:multi}. We further compare our results to previous findings in Section \ref{disc:lit}. We discuss implications for all of these results and the impact on our understanding of SN progenitors in Section \ref{sec:disc}.

\subsection{Individual Stellar Population Properties}
\label{sec:indiv}
We first compare redshift distributions across our spectroscopic SN samples. We plot the redshift histograms of each SN class in Figure \ref{fig:compare} and list the population medians and 68\% confidence intervals in Table \ref{tab:sp_results}. It appears that the redshift distributions of SNe Ia, Ib/c, II, and IIn are roughly consistent with each other. However, the distributions of SNe Ia and IIn veer toward higher redshifts than the Ib/c and II populations. As SNe Ia and IIn are generally brighter than Ib/c and II \citep{perley2020,ams+2023}, this difference in redshift is easily explained by an observational bias in our spectroscopic sample. We note that a majority of our spectroscopic sample comes from BTS (Section \ref{sec:sample}), which attempts spectroscopic classification for all SNe with peak magnitude $<19$ mag. This cut limits the detection of SNe Ib/c and II beyond $z\gtrsim 0.05$ \citep{Fremling2020}.

We next explore the stellar masses, sSFRs, stellar population ages, $Z_*$, and $A_V$ of our SN host samples. We highlight CDFs of these properties in Figure \ref{fig:compare} and note the population medians and 68\% confidence intervals in Table \ref{tab:sp_results}. We start by comparing the stellar population properties of SNe Ia hosts to those of the CCSNe populations. We find that SNe Ia tend to occur in galaxies with higher stellar mass and lower sSFR than CCSNe. We verify this by performing 2500 AD tests each (achieved in the same way as described in Section \ref{sec:mass}) between the stellar mass and sSFR distributions of SNe Ia to that of SNe Ib/c, II, and IIn. In all tests between stellar mass distributions, $P_{AD}<0.05$, strongly suggesting that SNe Ia prefer higher mass galaxies. The sSFRs of SNe Ia vs CCSN hosts also statistically differ: 100\% of $P_{AD}<0.05$ when comparing the sSFRs of SNe Ia to SNe II and IIn, and 88\% of $P_{AD}<0.05$ when comparing to SNe Ib/c. Moreover, SNe Ia appear to occur in older galaxies with higher $Z_*$ than CCSNe. Once again, the trend in stellar population ages is confirmed with statistical significance, as 55\%, 100\%, and 90\% of AD tests result in $P_{AD}<0.05$ when comparing SNe Ia host stellar population ages to those of SNe Ib/c, II, and IIn, respectively. We are only able to reject the null hypothesis that SNe Ia trace the same $Z_*$ distribution as SNe II and IIn (97\% and 70\%  of $P_{AD}<0.05$, respectively), as only 5\% of $P_{AD}<0.05$ when comparing $Z_*$ between SNe Ia and Ib/c. Overall, we find strong statistical support that SNe Ia occur in older, more massive and metal-rich galaxies with lower sSFRs than CCSNe. 

We next seek to contrast the host properties of SNe Ib/c to SNe II: two CCSN subtypes whose presumed progenitors undergo different massive star evolutions (see Section \ref{sec:prog}). Visually, our sample of SN Ib/c hosts appears to be more massive, metal-rich and less star-forming compared to SN II hosts. When performing AD tests between SN Ib/c and II host stellar population properties, we find that we are only able to reject the null hypothesis that they trace the same underlying distributions in stellar mass and sSFR (77\% of $P_{AD}<0.05$ for stellar mass and 58\% of $P_{AD}<0.05$ for sSFR). We cannot confirm the visual trend that SNe Ib/c trace more metal-rich hosts than SNe II with statistical significance: only 5\% of $P_{AD}<0.05$. We further note that their SFR distributions are very similar: only 2\% of $P_{AD}<0.05$. We interpret these results in the context of their progenitors in Section \ref{sec:prog}.

Lastly, we compare the host properties of SNe II(P/L) and IIn. We do not find any statistical support that SNe II and IIn trace different stellar mass, sSFR, age, or $Z_*$ distributions, as only a small fraction of AD tests between the host properties yield $P_{AD}<0.05$ (36\% for stellar mass, 5\% for sSFR, 2\% for age, and 6\% for $Z_*$). We do, however, find that SNe IIn hosts are dustier than SNe II hosts. We are able to reject 75\% of AD tests when comparing the host $A_V$ distributions of SNe II and IIn. We further find that SNe IIn hosts trace higher $A_V$ than Ib/c hosts (55\% of AD tests result in $P_{AD}<0.05$), suggesting their hosts are dustier than most CCSNe types. On the other hand, SNe IIn hosts have a more similar $A_V$ distribution to SNe Ia hosts: $<1\%$ of $P_{AD}<0.05$ when performing an AD test between their hosts' $A_V$.

\subsection{Multivariate SFR-$M_*$ Analysis}
\label{sec:multi}
Given that host galaxies of specific transient populations are characterized by a combination of stellar population properties, we next explore how our SN hosts track a multivariate galaxy-phase space. Specifically, we seek to understand how mass or SFR-weighted our host populations are by analyzing them in a combined SFR-$M_*$ plane. The purpose of this is twofold. For one, how dependent transient populations are on SFR and $M_*$ reveals the nature of their progenitors; i.e., we may expect that transients with very young progenitors will exclusively depend on SFR, while transients with older, compact object progenitors may be more closely linked with $M_*$. Additionally, this analysis will serve as a basis for how these different SN populations track stellar mass and SFR in the Universe, which will be useful for developing pure samples of these populations in the Rubin era. Here, we follow methods developed in \citet{hm2025}\footnote{https://github.com/hoasaf3/host-galaxies-stats}, where they derive a multivariate rejection test to determine if weighted combinations of SFR and mass can be ruled out to describe the SFR-$M_*$ distribution of an observed host galaxy population.

Following the methodology in \citet{hm2025}, we begin by sampling mock galaxy populations from a weighted SFR-$M_*$ distribution. The ``unweighted" distribution comes from \citet{leja2022}, wherein they derive a SFR-$M_*$($z$) probability density distribution, $\rho (M_*, SFR | z)$, for observed galaxies in the COSMOS2015 ($0.2 < z < 0.8$; \citealt{Laigle2016}) and 3D-Hubble Space Telescope (3D-HST; $z > 0.5$; \citealt{Skelton2014}) galaxy surveys. \citet{leja2022} models the stellar population properties of this galaxy population with \texttt{Prospector} and uses a non-parametric SFH; thus, we expect that the derived properties are complementary to the ones inferred here. We note that the \citet{leja2022} $\rho (M_*, SFR | z$) only includes galaxies with redshifts $0.2 < z < 3$ and $M_* > 10^8 M_\odot$. To make this function more practical for transient populations, \citet{hm2025} extrapolates this distribution to $z< 0.2$ and $M_* = 10^7 M_\odot$. We refer the reader to their text for more details on this extrapolation. 

A weighted SFR-$M_*$ distribution takes the form: $W \times \rho (M_*, SFR | z)$, where the $i^\textrm{th}$ galaxy in a sample has weight:
\begin{equation}
\label{eq:weights}
    W_i = A \Big( \frac{M_{*,i}}{M_\odot}\Big) + B \Big( \frac{\textrm{SFR}_i}{M_\odot \textrm{yr}^{-1}}\Big).
\end{equation}
If $A=0$, the function reduces to an SFR-weighted distribution, and if $B=0$, the function reduces to a mass-weighted distribution. We create 30 weighted distributions from $-15 \leq \log(A/B) \leq -6$, where $A+B=1$, spanning a broad range of mass and SFR weights. We then sample 32,000 galaxies at each of the redshifts within a specific host population for the 30 weighted distributions. Within both our SN host sample and the mock galaxy sample, we only include galaxies with $z \leq 0.2$, to mitigate effects of galaxy evolution that increase as we expand to higher $z$. Indeed, \citet{hm2025} also suggest analyzing results of their test within specific redshift bins, with spacing of $\approx0.2$. In any case, we do not have a substantial SN population above this limit (there are only 13 SNe total at $z>0.2$ in our spectroscopic sample), thus this restriction will not bias our results. We furthermore apply an optical bias to the weighted distributions so that we do not include any galaxies with optical magnitude $>23.5$~mag, similar to the observational bias of our SN host sample. 

\begin{figure}[t]
\centering
\includegraphics[width=0.45\textwidth]{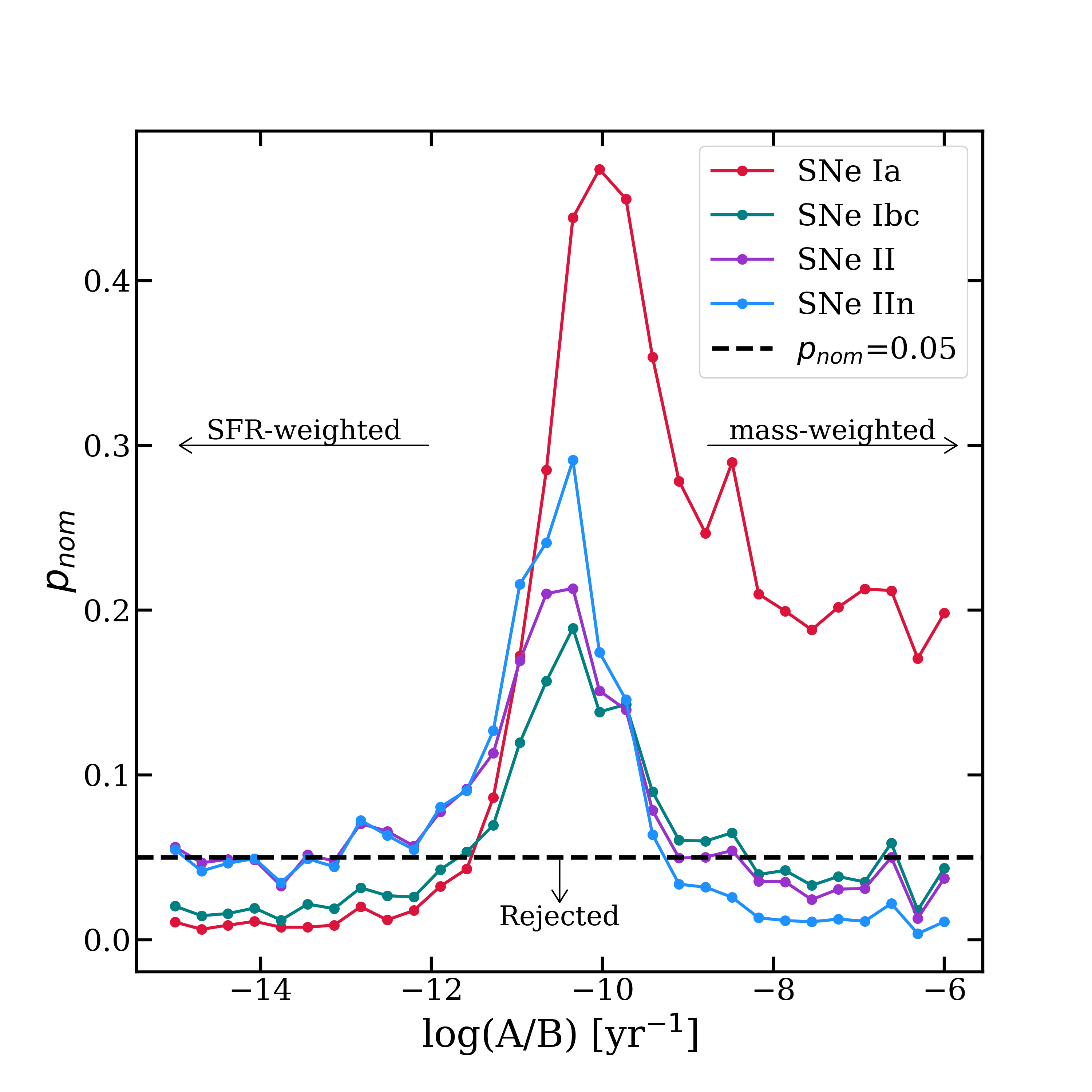}
\vspace{-0.1in}
\caption{Results of the \citet{hm2025} multivariate rejection test for each SN host population, which determines if the SFR-$M_*$ distributions of observed host samples are drawn from weighted distributions of SFR and $M_*$ of a simulated field galaxy population. Here, weights are determined through log($A/B$), where lower log($A/B$) values describe more SFR-weighted distributions and higher log($A/B$) describe more mass-weighted distributions. We reject that a weighted distribution describes a host population SFR-$M_*$ distribution well when $p_\textrm{nom} < 0.05$ (dashed black line). }
\label{fig:pnom}
\end{figure}

To determine if the SFRs and $M_*$ of an observed host population can be drawn from a specific weighted distribution, \citet{hm2025} compare the likelihood $\mathcal{L}$ for both a weighted-mock galaxy dataset ($\mathcal{L_\textrm{mock}}$) and the observed host population ($\mathcal{L_\textrm{data}}$), given by:
\begin{equation}
   \mathcal{L} = \sum_{i=1}^{N} \ln [W_i \rho (M_*, SFR | z)].
\end{equation}
We only calculate $\mathcal{L_\textrm{data}}$ at the median SFRs and $M_*$ of our hosts. If the weighted distribution describes the observed host SFR-$M_*$ distribution well, we expect that $\mathcal{L_\textrm{data}}$ will be similar to $\mathcal{L_\textrm{mock}}$. To better quantify whether a specific weighted distribution can be rejected to describe an observed host population, \citet{hm2025} derive a nominal $p$-value, $p_\textrm{nom} = P (\mathcal{L}_\textrm{mock} \leq \mathcal{L}_\textrm{data} | \textrm{model})$, where P is a cumulative distribution. If $p_\textrm{nom} < 0.05$, it implies that there is only a 5\% chance that the weighted mock galaxy population has a likelihood as extreme as the observed SN host sample. Similar to other rejection tests, we reject the null hypothesis that the observed galaxy population is derived from the weighted model distribution when $p_\textrm{nom} < 0.05$. 

\begin{figure*}[t]
\centering
\includegraphics[width=0.9\textwidth]{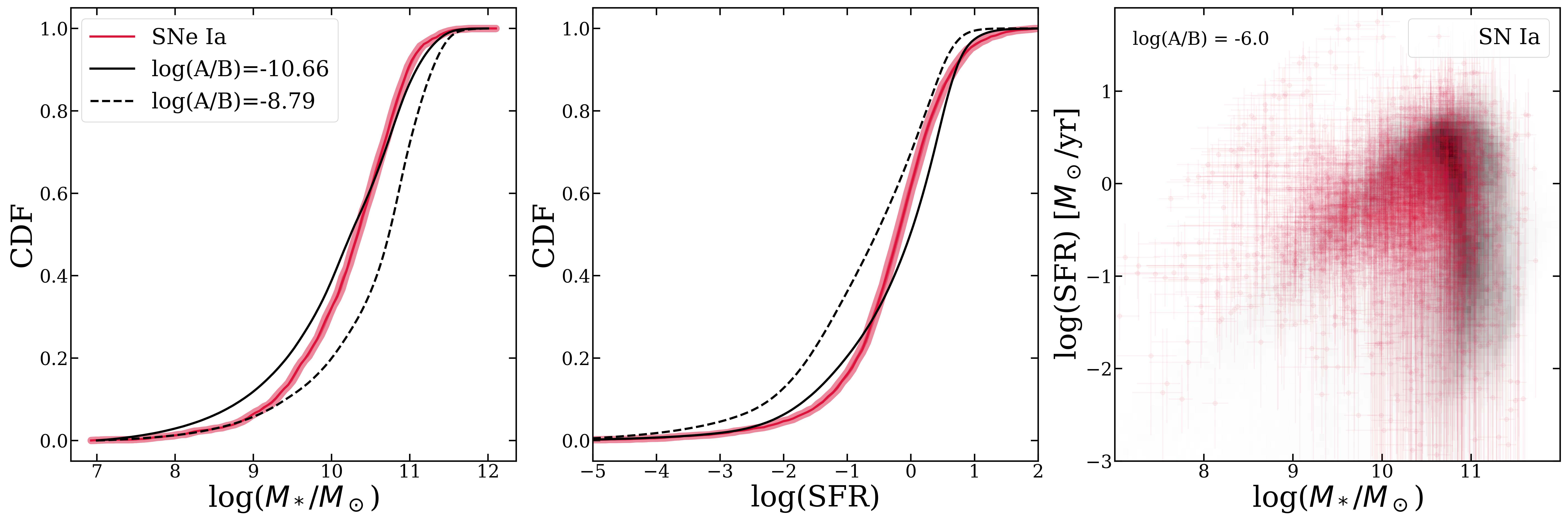}
\includegraphics[width=0.90\textwidth]{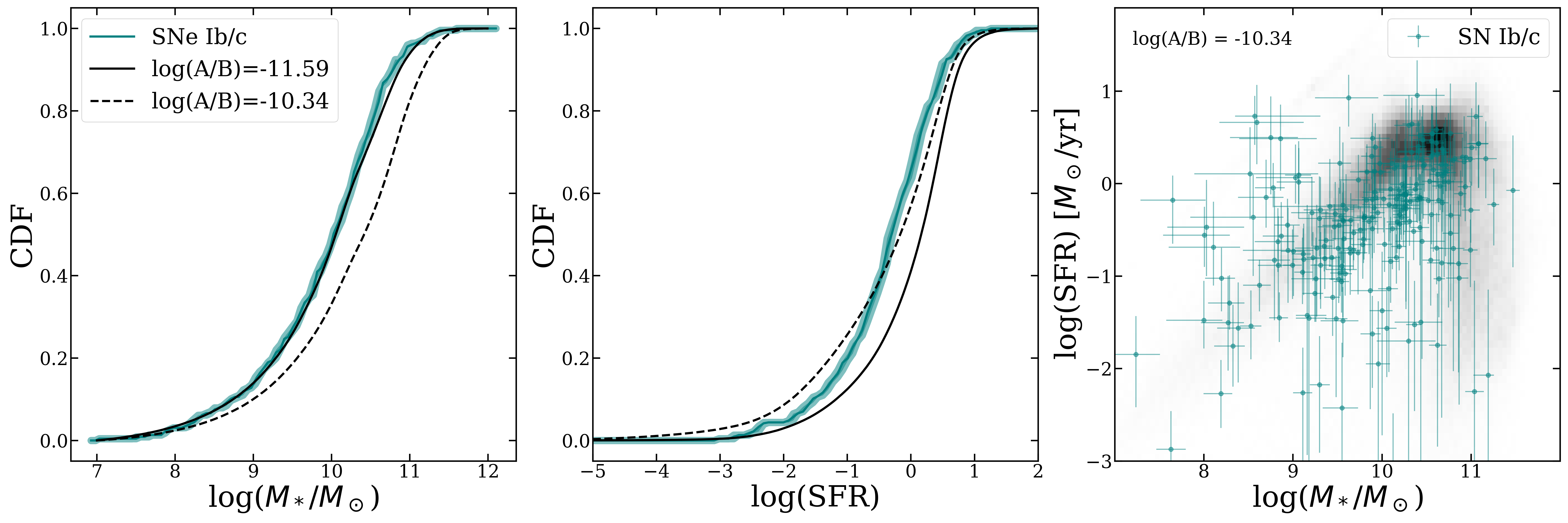}
\caption{CDFs in stellar mass (left column) and SFR (center column) for SNe Ia (red), Ib/c (green), II (purple), and IIn (blue). The straight black lines show the stellar mass and SFR CDFs of the field galaxy population drawn from a redshift-matched weighted distribution that fit the SN host stellar mass distribution well. Lower values of log($A/B$) represent more SFR-weighted distributions and higher values of log($A/B$) represent more mass-weighted distributions. For SNe Ib/c and IIn, these distributions are statistically supported. For SNe Ia and II, these distributions are chosen from visual inspection. The dashed black line shows the CDFs drawn from a more mass-weighted distribution that cannot be ruled out to fit the host SFR-$M_*$ distributions. In some cases, this results in CDFs which visually are a better to the observed SFR distributions than the straight black line. In the right column, we show an example weighted SFR-$M_*$ distribution (black histogram), compared to the inferred SFR and $M_*$ of each host population. 
\label{fig:weights}}
\end{figure*}

\addtocounter{figure}{-1}
\renewcommand{\thefigure}{\arabic{figure} (Cont.)}

\begin{figure*}
\centering
\includegraphics[width=0.9\textwidth]{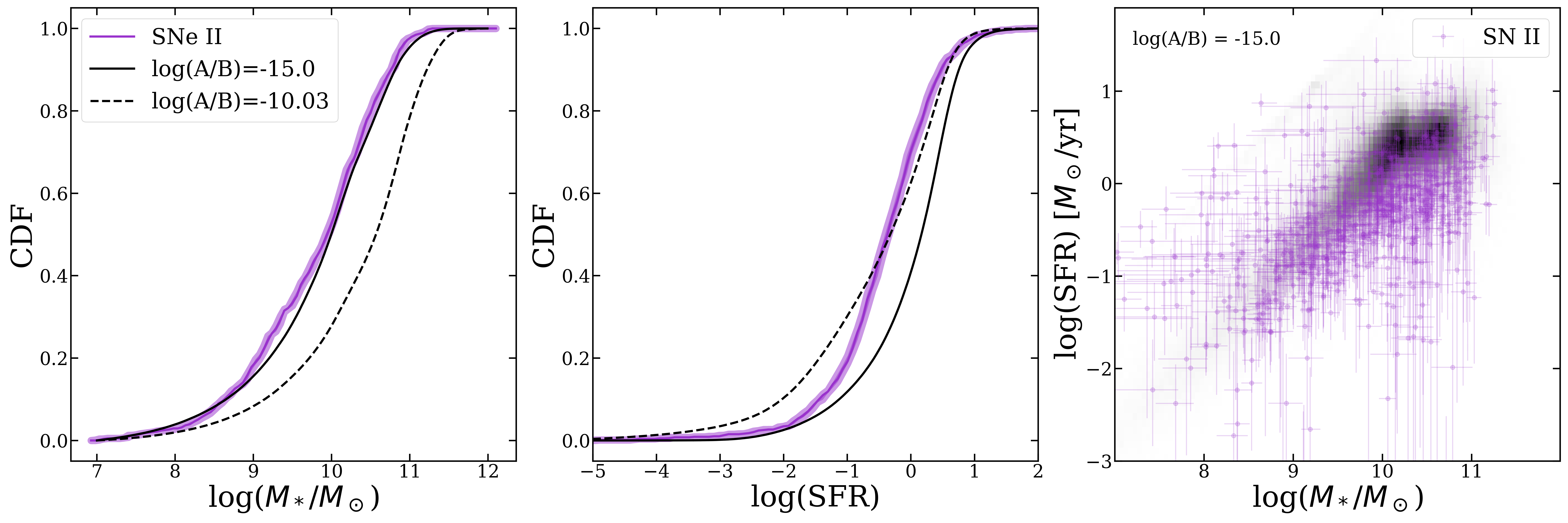}
\includegraphics[width=0.90\textwidth]{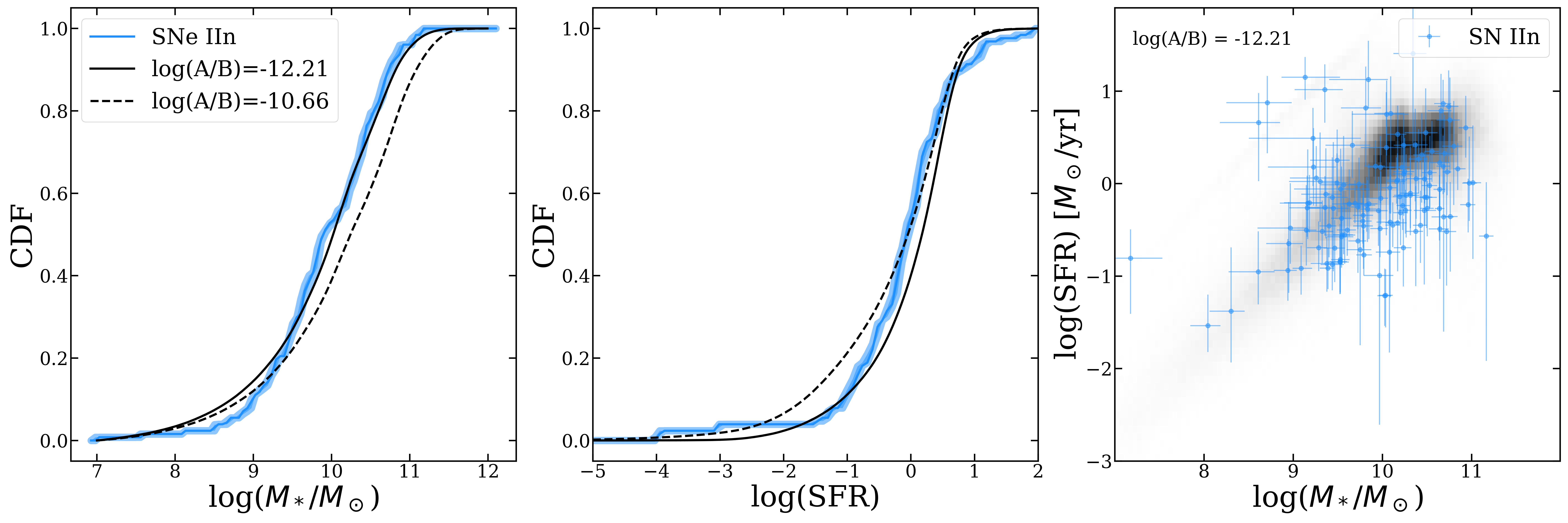}
\caption{CDFs in stellar mass (left column) and SFR (center column) for SNe Ia (red), Ib/c (green), II (purple), and IIn (blue). The straight black lines show the stellar mass and SFR CDFs of the field galaxy population drawn from a redshift-matched weighted distribution that fit the SN host stellar mass distribution well. Lower values of log($A/B$) represent more SFR-weighted distributions and higher values of log($A/B$) represent more mass-weighted distributions. For SNe Ib/c and IIn, these distributions are statistically supported. For SNe Ia and II, these distributions are chosen from visual inspection. The dashed black line shows the CDFs drawn from a more mass-weighted distribution that cannot be ruled out to fit the host SFR-$M_*$ distributions. In some cases, this results in CDFs which visually are a better to the observed SFR distributions than the straight black line. In the right column, we show an example weighted SFR-$M_*$ distribution (black histogram), compared to the inferred SFR and $M_*$ of each host population.}
\end{figure*}

\addtocounter{figure}{-1}
\renewcommand{\thefigure}{\arabic{figure} (Cont.)}

We show $p_\textrm{nom}$ derived at each log($A/B$) for all SN host samples in Figure \ref{fig:pnom}. For SNe Ia, we clearly rule out the most SFR-weighted distributions, but cannot rule out the most mass-weighted distributions. We find that we are able to rule out the most extreme mass and SFR-weighted distributions for SNe Ib/c, with their hosts initially appearing to prefer a more neutral combination of weights. Finally, we see that we cannot rule out some of the most SFR-weighted distributions for SNe II and IIn, but we are able to rule out the most mass-weighted distributions.

To better understand if the weighted distributions that cannot be ruled out describe our host populations well, we compare their 1D distributions of SFR and $M_*$ to those of our SN host samples. As done in \citet{hm2025}, we perform Kolmogorov-Smirnov (KS) tests to compare these distributions. To encapsulate the uncertainty on the result of the KS test, we build 2500 distributions in SFR and $M_*$ for each SN host population, using all \texttt{SBI++} chains for each host, and perform a KS tests between the weighted distributions and each of the 2500 realizations. We determine the median $p$-value ($\langle p_\textrm{KS} \rangle$) for all 2500 tests, and reject the null hypothesis that the SN host SFR or $M_*$ distribution traces the respective distribution of the weighted galaxy sample if the $\langle p_\textrm{KS} \rangle < 0.05$. We use the same redshift-matched weighted galaxy samples as discussed previously for these tests. 

We do not find any weighted SFR distributions that result in $\langle p_\textrm{KS} \rangle > 0.05$ for any SN population. For SNe Ib/c and IIn, we find that only the KS tests between their $M_*$ distributions and the most SFR-weighted $M_*$ distributions that cannot be ruled out in the multivariate test result in $\langle p_\textrm{KS} \rangle > 0.05$. For SNe Ib/c, this refers to $M_*$ distributions drawn from the weighted galaxy samples where $-11.6 \lesssim \log(A/B) \lesssim -11.3$. For SNe IIn, these are the $M_*$ distributions where $-15 \lesssim \log(A/B) \lesssim -11.3$. We do not find any $M_*$ distributions that result in a $\langle p_\textrm{KS} \rangle > 0.05$ for SNe Ia and II. In Figure \ref{fig:weights}, we show example 1D weighted distributions in $M_*$ and SFR compared to our host population. We plot the distributions that resulted in the maximum $\langle p_\textrm{KS} \rangle$ for $M_*$ for all SN populations (black lines; even in the cases of SNe Ia and II, in which all tests were rejected) and a more mass-weighted distribution (dashed black lines). For SNe Ib/c, II, and IIn, the more mass-weighted distributions visually appear to better fit their SFR distributions. 

In general, we find that the more SFR-weighted distributions over-predict the SFRs of our host population and the more mass-weighted distributions over-predict their stellar masses. Given that we lack statistical support to better constrain the preferred model for each of our host population, we focus on qualitative comparisons of our SN hosts in the SFR-$M_*$ plane to various weighted SFR-$M_*$ distributions and each other\footnote{We do not attempt to find a ``best"-fit solution, as this would require a full evaluation and normalization of the likelihoods, which goes beyond the scope of this paper.}. We show example weighted distributions in the SFR-$M_*$ plane compared to our host samples in Figure \ref{fig:weights}. Visually, we see that the highest density of hosts in each of our SN host samples traces the star-forming main sequence (SFMS): a well-known, observed galaxy correlation followed by star-forming galaxies as they gain in stellar mass \citep{whitaker2014, speagle2014, leja2022}. SN Ia and Ib/c hosts also appear to have a high density ``tail" towards higher stellar mass, lower SFR galaxies, with SNe Ia appearing to have a higher density of hosts within this tail than SNe Ib/c. This tail does not appear to be as strong, or present at all, for SNe II and IIn hosts.

To better quantify the density of these tails,  we count the number of quiescent hosts, which typically occupy this high-$M_*$, low SFR space \citep{leja2022}, in each of our SN samples. Using the \citet{Tachella2021} formula to determine if a galaxy is quiescent from its $z$ and median sSFR, we find that 12\% of SNe Ia hosts, 9\% of SNe Ib/c hosts, 3\% of SNe II hosts, and only 1 SNe IIn host can be considered quiescent. For the CCSNe hosts, we confirm that the majority appear to be redder and/or elliptical galaxies, implying the presence of older and more evolved stellar populations, which verifies their characterization as quiescent. Thus, we confirm the visual inspection of the trends that SNe Ia and Ib/c have a higher density of high-$M_*$, low SFR galaxies than SNe II and IIn. 

When observing trends in the weighted field galaxy SFR-$M_*$ distributions (a variety of which are shown in Figure \ref{fig:weights}), it appears that as $\log(A/B) \rightarrow \infty$ (i.e., become more mass-weighted), this high-$M_*$, low SFR tail becomes stronger. Hence, because we find a higher density of SNe Ia and Ib/c hosts along this tail, we infer that their preferred distributions must be more mass-weighted than SNe II and IIn. Moreover, SNe Ia hosts likely follow a more mass-weighted distribution than SNe Ib/c hosts, given that SNe Ia have a higher density of massive, low SFR host galaxies (and more quiescent hosts) than SNe Ib/c.

While it may seem concerning that none of the weighted distributions perfectly capture the 1D SFR and $M_*$ distributions of our host populations, this probably is just suggesting that the form of the weighted model is not the most accurate description of these hosts population. Indeed, as implied in \citet{hm2025}, the relationship between SFR-$M_*$ may not linear, which was assumed here, or there may be other important parameters worth considering. For example, metallicity may play a larger role in shaping the SN host SFR-$M_*$ distributions than assumed in this work. Given that the $Z_*$ estimates inferred here are likely poorly constrained and we do not have the correct data to determine gas-phase metallicities, we do not attempt to include metallicity in our weighted distribution model. Moreover, it is possible that the weighted model does not properly consider all selection effects of our observed SN and host populations; we only have limited the weighted distributions by optical magnitude and there may be other important factors to consider like e.g., dust extinction \citep{fox2021, grayling2023, mme+2025}. We strongly encourage any future studies on the mass and SFR-weights of observed host populations to consider the role and impact of these effects.

\subsection{Comparisons to Previous Literature}
\label{disc:lit}
Finally, we make a brief note about comparing our findings to those in other studies. We begin by comparing our results to those in \citet{schulze2021}, which analyzes the host galaxies of $\approx 900$ CCSNe discovered by the Palomar Transient Factory (PTF; \citealt{law2009, rau2009}). We find that all stellar mass population medians for our CCSNe hosts are higher than those reported in \citet{schulze2021}, with population medians differing by $\gtrsim 0.3$~dex. There are several notable reasons for these differences. For one, the choice of SFH in the stellar population model strongly affects the inferred stellar mass. \citet{schulze2021} employs a delayed-$\tau$ SFH ($\text{SFH} \propto t*e^{-t/\tau}$, where $\tau$ is the e-folding timescale) within their \texttt{Prospector} nested sampling stellar population modeling. Stellar masses inferred from non-parametric SFHs, as done here, are known to be 25-100\% larger than those inferred with a delayed-$\tau$ SFH \citep{Leja2019}. We also would expect non-parametric SFHs to skew results to lower SFRs/sSFRs than a delayed-$\tau$ SFH, and this difference is apparent between the \citet{schulze2021} and our samples. Moreover, the \citet{schulze2021} SN host sample includes many more hosts with optical magnitude $>23.5$~mag (the limiting magnitude of our sample), which naturally would extend their host populations to include more lower mass galaxies.

While the population medians may differ between our and the \citet{schulze2021} SN host samples, we do recover the same general trend between SNe Ib/c and SNe II hosts, i.e., that SNe Ib/c hosts are more massive. In addition, when comparing stellar population property estimates of the same CCSNe hosts (we test \texttt{FrankenBlast} on five randomly selected hosts from their sample), we generally find that the results are in good agreement. Thus, we do not believe that the differences between in our CCSNe host stellar population property results are concerning as they can be explained easily by differences in the stellar population modeling technique and the sample selection. These differences do highlight an important caveat in host galaxy studies: we may expect that general trends between transient host populations will be robust across various studies; however, it may be difficult to compare results directly between studies if the hosts were modeled under different assumptions (as is often true across galaxy SED fitting codes) or if the samples have different selection effects. This further emphasizes the requirement of uniformity in host modeling when analyzing large populations.

We next turn our attention to the results in \citet{qz2024}, which studies the hosts of $\approx 17,000$ SNe Ia and CCSNe. Similar to our study, \citet{qz2024} determines how mass or SFR-weighted different SN host populations are by comparing the host multivariate SFR-$M_*$ distributions to weighted distributions of a redshift-matched, observed galaxy populations in the Sloan Digital Sky Survey (SDSS) Main Galaxy Sample \citep{strauss2002}. Their main findings are consistent with ours: SNe Ib/c hosts have more mass-weighted SFR-$M_*$ distributions than SNe II hosts, SNe II and IIn have no distinguishable differences in the SFR-$M_*$ plane, and SNe Ia host SFR-$M_*$ distributions are more mass-weighted than that of CCSNe hosts. The uniformity in conclusions across \citet{qz2024} and our studies not only verifies our findings, but also validates the \citet{hm2025} methods for determining how mass or SFR-weighted different transient host populations are.

\section{Discussion}
\label{sec:disc}

\subsection{SNe Progenitor Implications}
\label{sec:prog}
Based on our analysis of the host galaxy stellar population properties of our spectroscopic SN sample in Section \ref{sec:transientcompare}, here, we explore what our findings tell us about their progenitors. The general trend of SNe Ia towards more more massive, less star-forming, older, and more metal-rich galaxies than CCSNe is expected. The WD progenitor of SNe Ia is assumed to have a much longer delay time than the CCSNe massive star progenitor ($\approx 100$~Myr minimum compared to $\approx10$~Myr for CCSNe progenitors), which forces them to more regularly occur in more evolved galaxies \citep{mennekens2010, anderson2012, mm2012, maoz2014, chen2021, wiseman2021}. This is also verified in our multivariate SFR-$M_*$ test, where we clearly see that SNe Ia hosts are drawn from more mass-weighted solutions than CCSNe hosts. We further note that $\log(A/B)$ weights determined here have previously been used to describe the SNe Ia rate as a function of host SFR and $M_*$ and understand their progenitor delay time distribution (e.g, \citealt{sb2005, sullivan2006}). Our results are consistent with \citet{sb2005} and \citet{sullivan2006}, where they find $\log(A/B) = -10.76$~yr$^{-1}$ and $\log(A/B) = -9.86$~yr$^{-1}$, respectively. This more ``mixed" $\log(A/B)$ solution, where neither SFR nor $M_*$ is over-preferred, suggests that the SNe Ia progenitor population comprises both a prompt channel (i.e., one that is more connected to recent star formation) and a delayed channel (more connected to $M_*$), as has also been found in other works \citep{mannucci2005, mm2012, smith2012, wiseman2021}. We also show that we cannot reject the most mass-weighted solutions for SNe Ia ($B \rightarrow \infty$). If we find that these solutions still cannot be rejected with larger SNe Ia populations discovered through Rubin and \textit{Roman}, this may hint that SNe Ia have a larger population of longer delay time progenitors than previously known. In general, our conclusions with SNe Ia host populations are unsurprising.

We find the preference of SNe Ib/c on more massive, less star-forming hosts, with more mass-weighted SFR-$M_*$ solutions than SNe II more intriguing. We note that these trends have been confirmed previously \citep{arcavi2010, li2011, graur2017, schulze2021, qz2024}. At face value, we may infer from host properties alone than the SNe Ib/c progenitor may be more evolved than SNe II. However, this does not cohere to current progenitor models. There are two main progenitor theories for SNe Ib/c: either, they are solitaire Wolf-Rayet stars ($M_\textrm{ZAMS} \approx 20-25 M_\odot$; \citealt{smith2011, smith2014, smartt2015}) that became stripped due to strong winds \citep{conti1975}, or they originate in a close-binary system  \citep{smartt2009, eldridge2013, yoon2015, solar2024} and have their outer envelopes stripped through binary interactions, with the latter likely being the dominant channel. On the other hand, SNe II derive from massive stars ($M_\textrm{ZAMS} > 8 M_\odot$) that retain their outer hydrogen envelopes and the vast majority have red super giant (RSG) progenitors \citep{burrows1995, smartt2009,  goldberg2020, dessart2021}. If SNe Ib/c are derived from single star systems, then we would not expect that their delay times would be distinct from that of SNe II. On the other hand, SNe Ib/c may have slightly longer delay times if they were to derive from binary systems ($\approx 200$~Myr maximum for binaries compared to $\approx$tens Myr for single stars; \citealt{zapartas2017}). Yet, this timescale is not likely to be long enough to account for significant differences in host galaxy populations as galaxies evolve in stellar mass more slowly than this \citep{peng2010, torrey2018}. Thus, it is unlikely that the host galaxies of these CCSNe subtypes differ because the SNe Ib/c progenitor is more evolved. Moreover, \citet{anderson2012} argue that SNe Ib/c progenitors may have shorter delay times than SNe II, given their closer proximity to younger HII regions within their hosts than SNe II. Although we do not explore local populations here, this may give further insight on the nature and timescales of SNe Ib/c and II progenitors.

Our results could also indicate that SNe Ib/c prefer more evolved hosts for the right conditions for progenitor formation. More evolved, massive galaxies tend to be more metal-rich \citep{gcb+05, peng2015, trussler2020, trussler2021, bsb2024}. Previous studies have found that SNe Ib/c hosts have higher gas-phase metallicities than SNe II hosts, determined through spectroscopic surveys of their global host environments \citep{prieto2008, kelly2012, qz2024} and local HII regions near the SNe \citep{anderson2010, modjaz2011, galbany2016}, as well as higher stellar metallicities determined through host photometry \citep{schulze2021}. Although our results are not indicative of whether SNe Ib/c hosts are more metal-rich (our $Z_*$ measurements are likley not well-constrained and we do not attempt to estimate gas-phase metallicities), we proceed under the assumption that higher-$M_*$ galaxies are more evolved and have higher gas-phase and stellar metallicities. Higher metallicity may allow more single star systems to explode as stripped envelope CCSNe due to higher mass loss rates from metal-driven winds \citep{heger2003}; thus, if SNe Ib/c are derived from solitaire Wolf-Rayet stars, their preference for more evolved hosts may naturally be explained by this. There does not, however, appear to be any dependence on massive star close binary fractions with metallicity \citep{sana2025}. Although, we do note that the low-mass star close binary fraction significantly decreases with increasing metallicity \citep{badenes2018, moe2019, mazzola2020, niu2022}. Thus, if SNe Ib/c derive from binary progenitors, we may not expect this preference for more evolved hosts. On the other hand, it is possible that the SNe Ib/c class comprises both solitaire and binary progenitors and that these different progenitors may be associated with distinct environments, which should be explored further in the future. It will be interesting to see if these distinctions between SNe Ib/c and II hosts hold when transients are detected out to higher redshifts with Rubin and \textit{Roman}.

We next focus on the comparison between SNe II and IIn hosts. We find that they are only distinct in their global host $A_V$ distributions, as they even prefer similar SFR-weighted SFR-$M_*$ distributions. In general, it is assumed that SNe II and IIn will have similar host galaxies, as their respective RSG and assumed luminous blue variable (LBV; \citealt{galyam2007, smith2011}) massive star progenitors, respectively, likely have an equal dependence on star formation and stellar mass\footnote{We do not divide the SNe IIn class into luminous and ``normal" events as was done in \citet{schulze2021}, where it was shown that luminous SNe IIn may have a very slight preference for lower mass galaxies than SNe II.}. No preference for specific host properties has been previously shown between these classes \citep{schulze2021, qz2024}. The higher dust content in SNe IIn hosts, however, may be important for understanding their progenitor. The fraction of dust in a galaxy, which arises from star formation, SNe, and mass loss in asymptotic giant branch (AGB) stars \citep{dwek1998,tielens1998,edmunds2001,morgan2003,Conroy2013,zhukovska2014,dgv+2015,michalowski2015}, has been shown to be proportional to O/H gas-phase metallicity \citep{issa1990, schmidt1993, draine2007}. Several previous studies suggest that SNe IIn may trace more metal-rich galaxies or local environments than SNe II \citep{habergham2014,taddia2015,qz2024}. Thus, our finding that SNe IIn prefer dustier hosts than SNe II may be indicative of a gas-phase metallicity preference for their progenitors.

We note that higher $A_V$ in SNe IIn hosts could also be simply exposing an observational bias in our SN sample. We show in Section \ref{sec:indiv} that SNe IIn hosts not only have higher $A_V$ than SNe II, but also higher $A_V$ than SNe Ib/c, and a more similar $A_V$ distribution to SNe Ia. As SNe Ia and IIn are generally brighter than SNe Ib/c and II, we can infer that they are more likely to be to be detected in hosts with large global dust extinction than SNe Ib/c or II. Hence, this could be the reason that their host populations veer towards dustier galaxies. SNe Ib/c and II \textit{may} occur at similar rates in dusty galaxies, but they are just less likely to be detected given that the high host $A_V$ may overcome their SN light.

We finally contemplate the reasons that some CCSNe occur in quiescent galaxies, as we see that 9\% of SNe Ib/c hosts and 3\% of SNe II hosts can be considered quiescent. \citet{schulze2021} also reports that a small fraction of CCSNe (Ib/c, II, IIn) occur in galaxies with low amounts of active star formation: they find that 2\% of CCSNe hosts have sSFR $<10^{-11}$~yr$^{-1}$. We note that we find that a higher fraction of CCSNe hosts below this sSFR limit (9\%), however, our results are skewed to lower sSFRs than those in \citet{schulze2021} given our choice in SFH (see Section \ref{disc:lit}). Additionally, \citet{irani2022} and \citet{sedgwick2021} find that only $\approx 1$-$10\%$ of CCSNe occur in elliptical galaxies (a morphological characteristic), which are generally redder and tend to have lower amounts of active star formation than spiral galaxies \citep{trager2000, strateva2001, delucia2006, pxn+2023}. 

While certainly rare, we may expect CCSNe to occur in quiescent galaxies given that a majority still retain a non-negligible amount of star formation. Indeed, \citet{kaviraj2014} argue that elliptical galaxies account for $\approx14\%$ of Universe's star formation budget. An alternative theory is that their progenitors are not massive stars, but possibly low mass stars or WDs, which has been postulated for several SNe Ibn in quiescent galaxies \citep{sanders2013, hosseinzadeh2019}. However, it is difficult to explain the photometric and spectroscopic properties of SNe Ib/c and II events with a degenerate star model, making this explanation less favorable.  We thereby conclude that the most likely explanation for CCSNe in quiescent hosts is that there is enough residual star formation left in these hosts to form their massive star progenitors. Future studies focusing on the local environments of these CCSNe may uncover whether they are indeed occurring in regions with higher star formation than observed globally in their host.

\subsection{Rubin-Era Science}
\label{sec:lsst}
Here, we discuss how our work can be applied in the Rubin era, which is expected to release its first alerts in 2025 and survey data products in 2027. Importantly, Rubin not only promises to detect millions of transients each year, greatly expanding the population sizes of different SN and transient classes; but the survey depth will also allow detection of transients out to much higher redshifts. It is expected, for example, that SNe Ia and II will be observed out to $z\approx1.5$, SNe Ib/c out to $z\approx1$, and SNe IIn out to $z\approx2$ \citep{kessler2019}: significantly higher $z$ than either the spectroscopic or photometric SN samples studied here. \textit{Roman} will observe these same SN populations out to even higher redshifts \citep{hounsell2018, foley2019}. While this will allow us to better understand transient rates and environments at earlier cosmic times, it will also bring up multiple challenges. As we show in Section \ref{sec:specvphot}, the slight redshift differences between our spectroscopic and photometric SN populations do affect their stellar mass and sSFR distributions. This is further emphasized when comparing our results to those of \citet{wiseman2020}, in which higher redshift SNe Ia (mean $z=0.3$) and CCSNe (mean $z=0.14$) host stellar masses were determined. There, they find that a median SNe Ia host $M_* = 10^{10.15} M_\odot$ and CCSNe host $M_* = 10^{9.39} M_\odot$ ($>0.2$~dex lower than the medians for our SNe Ia and CCSNe populations), with $M_*$ distributions that skew towards more lower mass galaxies ($M_* < 10^9 M_\odot$ than our populations. These results strongly suggest that host populations within the same transient class will differ when expanding out to higher $z$, with more deviations expected between Rubin-era transients and currently studied host populations. Moreover, we currently do not have any baseline for what these SN host properties will be at higher-$z$; thus, it is difficult to predict how much these distributions will deviate. In addition, comparing host properties between transient populations will also become more challenging for the same reason. \citet{kessler2019} predicts a higher rate of SNe II than SNe Ib/c at higher redshift, purely due to predicted observability and current rate estimates. This may mean, for instance, that SNe II hosts will have higher sSFR and lower stellar mass than SNe Ib/c. Yet, this effect may not be due to their progenitors but rather the sample biases. Thus, our analysis of transient hosts in the Rubin era must be performed in a way that marginalizes over the nuances of redshift effects and other systemic biases that may effect our transient populations. Taking advantage of methods similar to the one proposed in \citet{hm2025}, where redshift is factored into multivariate SFR-$M_*$ distributions, will be crucial for these large population analyses. 

In preparation for Rubin-related science, we have built \texttt{SBI++} spec-$z$ and photo-$z$ models to determine stellar population properties of galaxies at $0\leq z \leq 2$ with Rubin $ugrizy$ filters. To properly noise-up the training set photometry, we acquire photometric SNR measurements from the \texttt{cosmoDC2} Rubin galaxy simulation \citep{cosmodc2}. Beyond the slightly increased redshift prior, the training set is created in the same way as discussed in Section \ref{sec:training}. These models are available for public use in the same location as all other \texttt{FrankenBlast} products.

\section{Conclusion}
\label{sec:conc}

In this paper, we develop a fast and reliable method to associate transients to their host galaxies, perform aperture photometry on the hosts, and determine their stellar population properties, including redshift, stellar mass, SFH and SFR, stellar metallicity, stellar population age, and dust extinction. These methods comprise \texttt{FrankenBlast}, based on the \texttt{Blast} web application.

We test \texttt{FrankenBlast} on 6676 spectroscopically classified SNe and 7756 photometrically classified SNe, the majority of which were discovered through YSE and ZTF. In total, we determine host galaxy stellar population properties for 9262 events. We first compare the redshifts, stellar masses, sSFRs, and host $A_V$ of spectroscopically-classified SN hosts to those of photometrically-classified events and list our main conclusions below:
\begin{itemize}
    \item We find that photometrically-classified SNe lie at higher redshifts than spectroscopically-classified SNe. While this makes sense, given that higher redshift transients are fainter and less likely to be observed spectroscopically than lower redshift transients, it also implies that that there will be differences in the distributions of other host properties that are redshift-dependent (e.g., stellar mass and sSFR). This effect will be crucial to consider in future studies with Rubin and \textit{Roman} transients, where transients will mostly be photometrically-classified and will occur at higher redshifts than current samples.
    \item We show that photometrically-classified SNe Ia, Ib/c, II, and IIn hosts trace higher stellar masses than the spectroscopically-classified host samples. When we restrict the photometrically-classifed host sample to only the events with high SN classification pseudo-probabilities and high host association probabilities (refined photometric sample), we find better agreement between the stellar mass distributions, but they generally still lean towards higher stellar masses. Thus, this difference in stellar mass is likely both a redshift effect (as lower mass galaxies are harder to detect at increasing redshift) and caused by contamination from misclassified events or mis-associated host galaxies.
    \item We find evidence that our refined photometric SN Ib/c sample is likely still impacted by misclassified SNe Ia, given that a handful of events with high stellar mass and sSFR have more host $A_V$ than observed for any of the spectroscopically-classified SNe Ib/c hosts. These properties are more similar to that of SNe Ia hosts and, indeed, the SN lightcurves and brightness of these events do appear more similar to those of SNe Ia, as well.
\end{itemize}

To better understand the progenitors of SNe Ia, Ib/c, II, and IIn, we analyze the host properties of their spectroscopically-classified samples. We list our main conclusions with this analysis below:
\begin{itemize}
    \item We find that SNe Ia hosts tend to be more massive and metal-rich, less star-forming, and older than the CCSNe (Ib/c, II, and IIn) hosts. We further determine that SNe Ia hosts prefer more mass-weighted solutions than CCSNe hosts to describe their multivariate SFR-$M_*$ distributions. This aligns well with current WD progenitor theories of SNe Ia and expectations of their longer delay times than massive star explosions.
    \item We also show that SNe Ib/c hosts are more massive, less star-forming and prefer more mass-weighted SFR-$M_*$ distributions than SNe II. If SNe Ib/c originate from solitary massive star progenitors, which more easily explode as stripped-envelope SNe in more evolved galaxies that are typically more metal-rich, than this preference is understood. This environmental dependency is more difficult to explain if SNe Ib/c derive from close-binary systems, as their formation is either not metallicity dependent or inversely dependent on metallicity (depending on progenitor star mass).
    \item We find that SNe Ia and IIn have higher global host $A_V$ than SNe Ib/c and II hosts, which may be an observational bias, as SNe Ib/c and II are fainter than SNe Ia and IIn.
    \item We recover a small fraction (9\%) of CCSNe in quiescent hosts. While unusual, this likely indicates that there is enough residual star formation in these hosts to produce massive stars that eventually undergo core-collapse.
\end{itemize}

The success of \texttt{FrankenBlast} in constraining host stellar population properties quickly ($\lesssim 15$~min) and keeping general SN host trends in alignment with many previous host studies, highlights that it will be a valuable tool for transient-related science in the Rubin and \textit{Roman} eras. The host properties extracted by \texttt{FrankenBlast}, especially redshift (which will be unknown for many of the Rubin-detected galaxies), can easily be fed into photometric classifiers for better inference on transient class (e.g., \citealt{boesky2025}) or to help discover anomalous events. Moreover, these tools can be used for progenitor inference on large populations of various transients via their host properties. Indeed, with the multitude of transients we expect to detect with Rubin, we can more precisely infer differences in the host galaxies between transient classes and determine firmer constraints on how mass or SFR-weighted their populations are: crucial for understanding how transient progenitors evolve over cosmic time. Overall, reliable and rapid tools like \texttt{FrankenBlast} will likely be essential for progress in population-level transients in the next decade of ``Big Data".

\section*{Acknowledgments}
The Villar Astro Time Lab acknowledges support through the David and Lucile Packard Foundation, the Research Corporation for Scientific Advancement (through a Cottrell Fellowship), the National Science Foundation under AST-2433718, AST-2407922 and AST-2406110, as well as an Aramont Fellowship for Emerging Science Research. This work is supported by the National Science Foundation under Cooperative Agreement PHY-2019786 (the NSF AI Institute for Artificial Intelligence and Fundamental Interactions).  

D.O.J. acknowledges support from NSF grants AST-2407632, AST-2429450, and AST-2510993, NASA grant 80NSSC24M0023, and HST/JWST grants HST-GO-17128.028 and JWST-GO-05324.031, awarded by the Space Telescope Science Institute (STScI), which is operated by the Association of Universities for Research in Astronomy, Inc., for NASA, under contract NAS5-26555.

The computations in this paper were run on the FASRC Cannon cluster supported by the FAS Division of Science Research Computing Group at Harvard University.

The Pan-STARRS1 Surveys (PS1) and the PS1 public science archive have been made possible through contributions by the Institute for Astronomy, the University of Hawaii, the Pan-STARRS Project Office, the Max-Planck Society and its participating institutes, the Max Planck Institute for Astronomy, Heidelberg and the Max Planck Institute for Extraterrestrial Physics, Garching, The Johns Hopkins University, Durham University, the University of Edinburgh, the Queen's University Belfast, the Harvard-Smithsonian Center for Astrophysics, the Las Cumbres Observatory Global Telescope Network Incorporated, the National Central University of Taiwan, the Space Telescope Science Institute, the National Aeronautics and Space Administration under Grant No. NNX08AR22G issued through the Planetary Science Division of the NASA Science Mission Directorate, the National Science Foundation Grant No. AST-1238877, the University of Maryland, Eotvos Lorand University (ELTE), the Los Alamos National Laboratory, and the Gordon and Betty Moore Foundation.

The Legacy Surveys consist of three individual and complementary projects: the Dark Energy Camera Legacy Survey (DECaLS; Proposal ID 2014B-0404; PIs: David Schlegel and Arjun Dey), the Beijing-Arizona Sky Survey (BASS; NOAO Prop. ID 2015A-0801; PIs: Zhou Xu and Xiaohui Fan), and the Mayall z-band Legacy Survey (MzLS; Prop. ID 2016A-0453; PI: Arjun Dey). DECaLS, BASS and MzLS together include data obtained, respectively, at the Blanco telescope, Cerro Tololo Inter-American Observatory, NSF’s NOIRLab; the Bok telescope, Steward Observatory, University of Arizona; and the Mayall telescope, Kitt Peak National Observatory, NOIRLab. Pipeline processing and analyses of the data were supported by NOIRLab and the Lawrence Berkeley National Laboratory (LBNL). The Legacy Surveys project is honored to be permitted to conduct astronomical research on Iolkam Du’ag (Kitt Peak), a mountain with particular significance to the Tohono O’odham Nation.

NOIRLab is operated by the Association of Universities for Research in Astronomy (AURA) under a cooperative agreement with the National Science Foundation. LBNL is managed by the Regents of the University of California under contract to the U.S. Department of Energy.

This project used data obtained with the Dark Energy Camera (DECam), which was constructed by the Dark Energy Survey (DES) collaboration. Funding for the DES Projects has been provided by the U.S. Department of Energy, the U.S. National Science Foundation, the Ministry of Science and Education of Spain, the Science and Technology Facilities Council of the United Kingdom, the Higher Education Funding Council for England, the National Center for Supercomputing Applications at the University of Illinois at Urbana-Champaign, the Kavli Institute of Cosmological Physics at the University of Chicago, Center for Cosmology and Astro-Particle Physics at the Ohio State University, the Mitchell Institute for Fundamental Physics and Astronomy at Texas A \& M University, Financiadora de Estudos e Projetos, Fundacao Carlos Chagas Filho de Amparo, Financiadora de Estudos e Projetos, Fundacao Carlos Chagas Filho de Amparo a Pesquisa do Estado do Rio de Janeiro, Conselho Nacional de Desenvolvimento Cientifico e Tecnologico and the Ministerio da Ciencia, Tecnologia e Inovacao, the Deutsche Forschungsgemeinschaft and the Collaborating Institutions in the Dark Energy Survey. The Collaborating Institutions are Argonne National Laboratory, the University of California at Santa Cruz, the University of Cambridge, Centro de Investigaciones Energeticas, Medioambientales y Tecnologicas-Madrid, the University of Chicago, University College London, the DES-Brazil Consortium, the University of Edinburgh, the Eidgenossische Technische Hochschule (ETH) Zurich, Fermi National Accelerator Laboratory, the University of Illinois at Urbana-Champaign, the Institut de Ciencies de l’Espai (IEEC/CSIC), the Institut de Fisica d’Altes Energies, Lawrence Berkeley National Laboratory, the Ludwig Maximilians Universitat Munchen and the associated Excellence Cluster Universe, the University of Michigan, NSF’s NOIRLab, the University of Nottingham, the Ohio State University, the University of Pennsylvania, the University of Portsmouth, SLAC National Accelerator Laboratory, Stanford University, the University of Sussex, and Texas A \& M University.

BASS is a key project of the Telescope Access Program (TAP), which has been funded by the National Astronomical Observatories of China, the Chinese Academy of Sciences (the Strategic Priority Research Program “The Emergence of Cosmological Structures” Grant  XDB09000000), and the Special Fund for Astronomy from the Ministry of Finance. The BASS is also supported by the External Cooperation Program of Chinese Academy of Sciences (Grant 114A11KYSB20160057), and Chinese National Natural Science Foundation (Grant 12120101003, 11433005).

The Legacy Survey team makes use of data products from the Near-Earth Object Wide-field Infrared Survey Explorer (NEOWISE), which is a project of the Jet Propulsion Laboratory/California Institute of Technology. NEOWISE is funded by the National Aeronautics and Space Administration.

The Legacy Surveys imaging of the DESI footprint is supported by the Director, Office of Science, Office of High Energy Physics of the U.S. Department of Energy under Contract No. DE-AC02-05CH1123, by the National Energy Research Scientific Computing Center, a DOE Office of Science User Facility under the same contract; and by the U.S. National Science Foundation, Division of Astronomical Sciences under Contract No. AST-0950945 to NOAO.

The Photometric Redshifts for the Legacy Surveys (PRLS) catalog used in this paper was produced thanks to funding from the U.S. Department of Energy Office of Science, Office of High Energy Physics via grant DE-SC0007914.

\vspace{5mm}
\software{\texttt{SBI++} \citep{wlv+2023}, \texttt{Pröst}, \texttt{Prospector} \citep{jlc+2021}, \texttt{Python-fsps} \citep{FSPS_2009, FSPS_2010}, \texttt{Dynesty} \citep{Dynesty}, \texttt{Astropy}}

\bibliography{refs}

\appendix 
\restartappendixnumbering

\section{SBI++ Comparison to Prospector}
\label{app:prosp_sbi}
Here, we highlight the differences between the \texttt{SBI++} and \texttt{Prospector} nested sampling SED fits of 100 randomly selected SN hosts within our spectroscopic sample. In Figure \ref{fig:zfix_p_s}, we show comparisons between \texttt{SBI++} spec-$z$ and \texttt{Prospector} nested sampling fixed-$z$ fits. In Figure \ref{fig:zfree_p_s}, we show how the \texttt{SBI++} photo-$z$ fits and \texttt{Prospector} nested sampling fits (with redshift set as a free parameter) compare to the \texttt{SBI++} spec-$z$ fits.

Finally, we determine how well-calibrated the \texttt{SBI++} and \texttt{Prospector} nested sampling posteriors on redshift are. For each \texttt{Prospector} nested sampling and \texttt{SBI++} photo-$z$ fit, we determine if the true redshift falls within the 10-90\% credible interval of the redshift posteriors (with a step-size of 10\%). We then build a CDF from these counts and show this in Figure \ref{fig:coverage}. If the posteriors are perfectly calibrated, then we would expect the CDFs to follow the 1:1 line. We find that both the CDFs generated from the \texttt{Prospector} nested sampling fits and the \texttt{SBI++} photo-$z$ fits fall below this line, suggesting that both methods are under-confident in their redshift determination. However, it is also clear that \texttt{SBI++} redshifts posteriors are more likely to capture the true redshift than \texttt{Prospector} nested sampling, given its CDF lies closer to the 1:1 line, highlighting that \texttt{SBI++} posteriors are indeed better calibrated.

\begin{figure*}[ht!]
\centering
\includegraphics[width=1.0\textwidth]{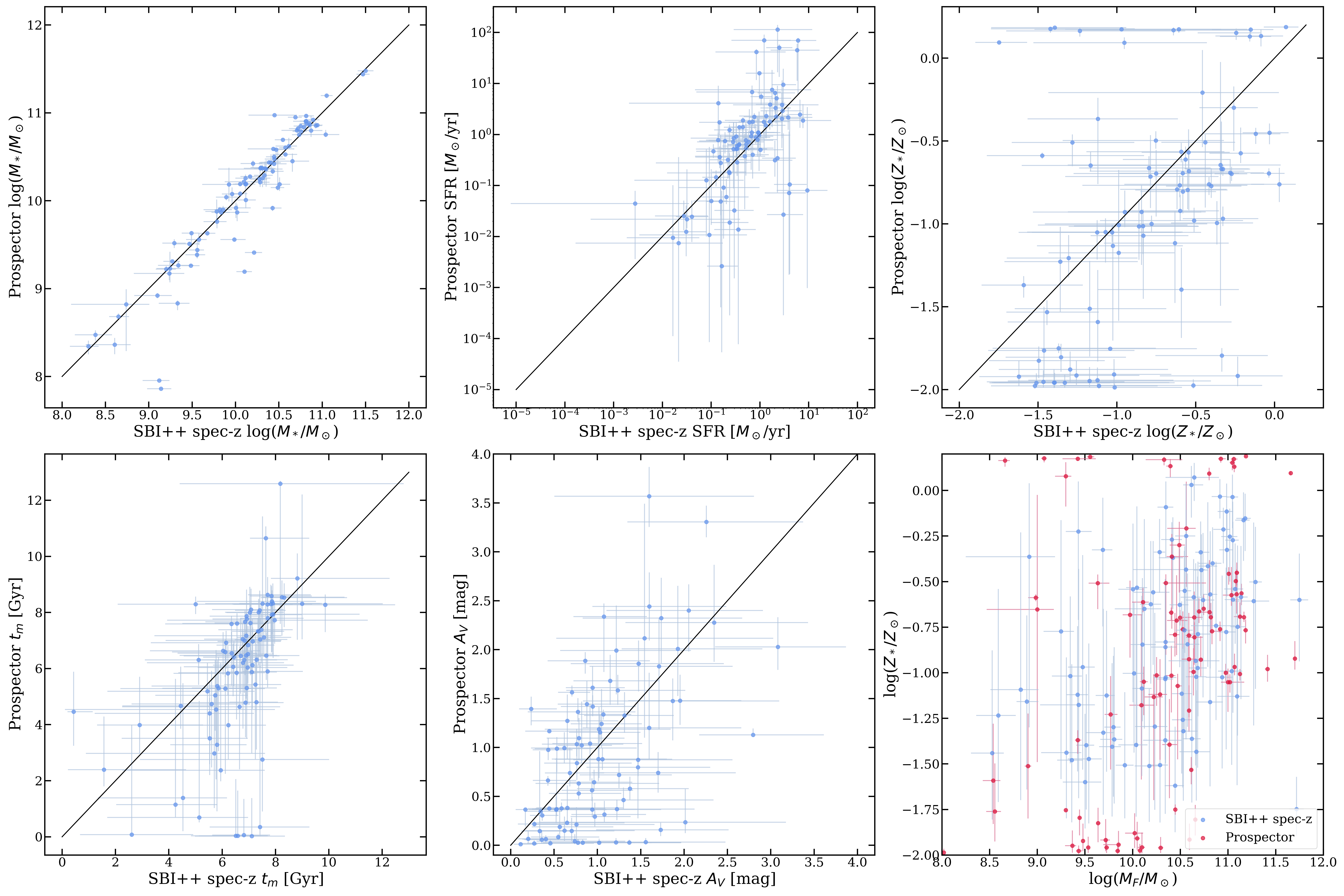}
\vspace{-0.1in}
\caption{Comparison of \texttt{SBI++} spec-$z$ fits to \texttt{Prospector} nested sampling fixed redshift fits for 100 randomly selected SN hosts in our spectroscopically-classified sample. The first five plots show the comparisons in stellar mass ($M_*$), SFR, metallicity ($Z_*$), mass-weighted age ($t_m$), and dust extinction ($A_V$). The black line designates the 1:1 ratio. The final plot (bottom right) shows $M_F$ and $Z_*$ inferred from \texttt{SBI++} (blue) and \texttt{Prospector} nested sampling (red). We generally find good agreement between the inferred stellar population properties. However, we note that some $t_m$, $Z_*$, and $A_V$ are substantially different, likely due to a failure in the \texttt{Prospector} nested sampling model fits. We further find that while the \texttt{SBI++} posteriors generally increase in $Z_*$ with increasing $M_F$, as expected from the $M_F$-$Z_*$ prior used to create the training set, a handful of \texttt{Prospector} posteriors are inconsistent with this relation, also likely due to a failure in the \texttt{Prospector} fit.}
\label{fig:zfix_p_s}
\end{figure*}

\begin{figure*}
\centering
\includegraphics[width=1.0\textwidth]{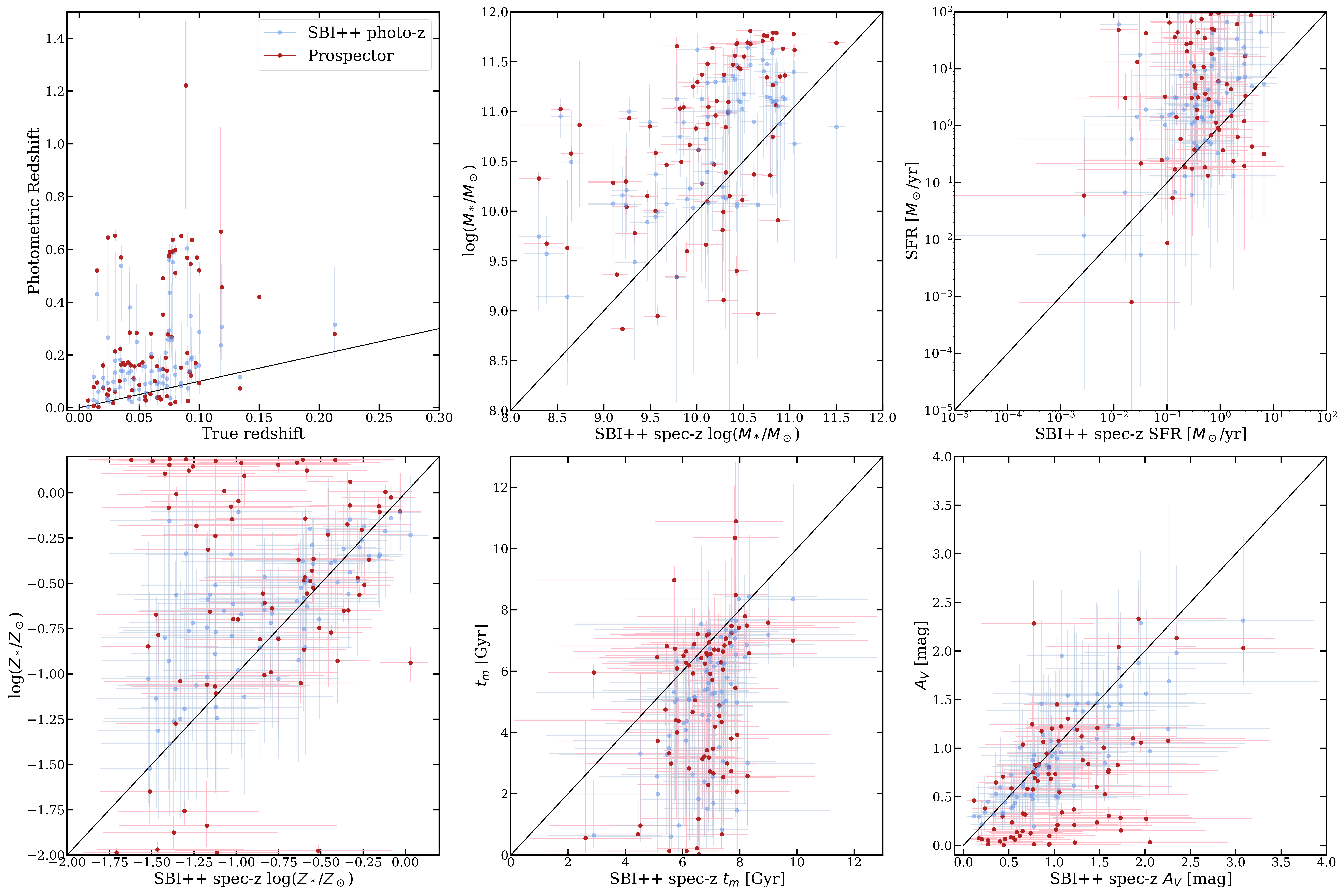}
\vspace{-0.1in}
\caption{Comparison of \texttt{SBI++} photo-$z$ inferred stellar population properties (blue) to \texttt{Prospector} nested sampling redshift-free stellar population properties (red) for 100 randomly selected SN hosts in our spectroscopically-classified sample. The black line in all plots shows the 1:1 ratio between these fits and the properties inferred from the \texttt{SBI++} spec-$z$ fits. We find that \texttt{Prospector} tends to produce too stringent posteriors on redshift that are unreliable, leading to incorrect inference on other properties. The \texttt{SBI++} photo-$z$ model, however, produces larger, more realistic uncertainties, that tend to capture the true value.}
\label{fig:zfree_p_s}
\end{figure*}

\begin{figure*}
\centering
\includegraphics[width=0.5\textwidth]{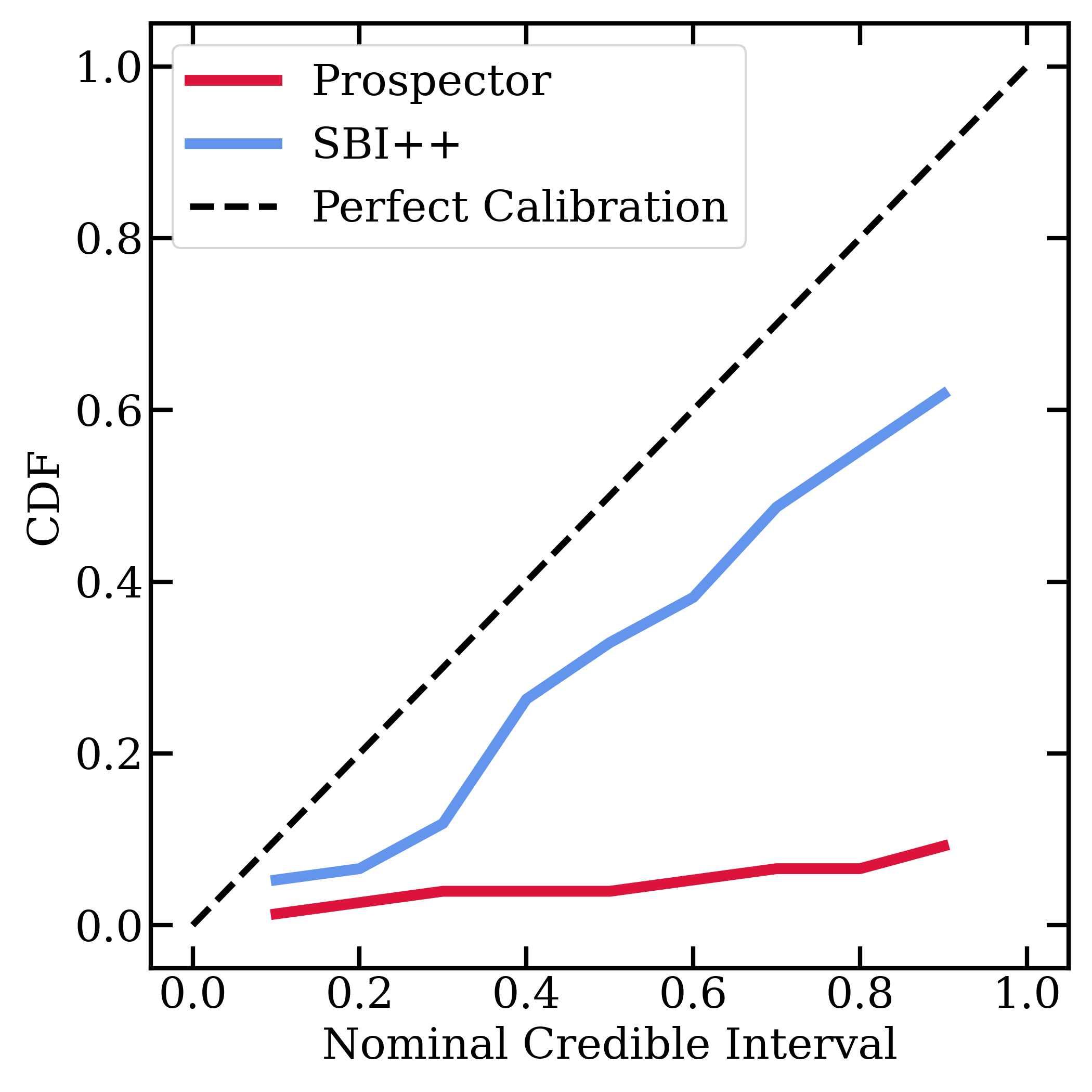}
\vspace{-0.1in}
\caption{CDFs built by determining if the true redshift of 100 randomly selected hosts fall within the 10-90\% credible interval of redshift posteriors generated from \texttt{SBI++} photo-$z$ fits (blue) and \texttt{Prospector} nested sampling fits (red). While neither method is perfectly calibrated (dashed black line), we find that \texttt{SBI++} posteriors are better calibrated than \texttt{Prospector} nested sampling, given that its CDF falls closer to the 1:1 line.}
\label{fig:coverage}
\end{figure*}

\end{document}